\documentclass[letterpaper,twocolumn,10pt]{article}
\usepackage{usenix}

\usepackage{arabtex}
\usepackage{utf8}

\usepackage[T1]{fontenc}
\usepackage{balance}
\usepackage{algorithm}
\usepackage{algpseudocode}
\usepackage{mathptmx}
\usepackage{amsmath}
\usepackage{wasysym}
\usepackage{colortbl}
\usepackage{longtable}

\usepackage{enumitem}
\usepackage{mathptmx}
\usepackage{graphicx}
\usepackage{array}
\usepackage{multirow}
\usepackage{booktabs}
\usepackage{rotating}
\usepackage{xparse}
\usepackage{tabularx}
\usepackage{url}
\usepackage{cite}
\usepackage{amssymb}
\usepackage{pifont}
\usepackage{subfig}
\usepackage{authblk}
\usepackage[font={small}]{caption}
\usepackage{titling}
\usepackage{makecell}
\usepackage{array}
\usepackage{hyperref}
\microtypecontext{spacing=nonfrench}
\newcolumntype{Y}{>{\centering\arraybackslash}X}
\newcolumntype{P}[1]{>{\centering\arraybackslash}p{#1}}

\usepackage{titlesec}
\titleformat*{\subsection}{\normalfont\fontsize{11}{12.5}\bfseries} 

\makeatletter
\renewcommand{\paragraph}{%
  \@startsection{paragraph}{4}%
  {\z@}{1.75ex \@plus 0.35ex \@minus 0.2ex}{-1em}%
  {\normalfont\normalsize\bfseries}%
}
\makeatother

\newcommand{\comment}[1]{\relax}
\newcommand{\eg}{e.g.\@\xspace}
\newcommand{\ie}{i.e.\@\xspace}
\newcommand{\etal}{et~al.\@\xspace}
\newcommand{\etc}{{\em etc.}\xspace}

\usepackage{soul}
\usepackage{xspace}
\usepackage{arydshln}
\definecolor{lightgray}{gray}{0.9}

\newcommand{\Merit}{\textit{Merit}\xspace}

\colorlet{pastelred}{red!7}
\sethlcolor{pastelred}

\usepackage[strings]{underscore}
\usepackage{adjustbox}
\usepackage{array}

\definecolor{armygreen}{rgb}{0.0, 0.5, 0.0}
\definecolor{light-gray}{gray}{0.9}

\newcommand{\myparagraph}[1]{\smallskip\textbf{#1}\hspace{3pt}}

\newlength{\qswidth}
\newlength{\nswidth}

\newlength{\bibitemsep}
\newlength{\bibparskip}
\let\oldthebibliography\thebibliography
\renewcommand\thebibliography[1]{%
  \oldthebibliography{#1}%
  \setlength{\parskip}{\bibitemsep}%
  \setlength{\itemsep}{\bibparskip}%
}
\usepackage{xcolor, soul}
\sethlcolor{yellow}
\begin{document}

\title{\Large \bf {OpenVPN is Open to VPN Fingerprinting}}

\author{
\vspace{-2ex}
    {Diwen Xue}$^{*}$ \quad {Reethika Ramesh}$^{*}$ \quad  {Arham Jain}$^{*}$ \quad {Michalis Kallitsis}$^\dagger$ \\\vspace{-2ex}
    {J. Alex Halderman}$^{*}$ \hspace{1em} {Jedidiah R. Crandall}$^\ddagger$ \hspace{1em} {Roya Ensafi}$^{*}$ \\
    \\
      $^{*}${University of Michigan} \hspace{2em} $^\dagger${Merit Network, Inc.}  \\ $^\ddagger$ {Arizona State University/Breakpointing Bad} 
}

\date{\vspace{-5ex}}

\maketitle
\thispagestyle{empty}
\pagestyle{empty}
\subsection*{Abstract}  
VPN adoption has seen steady growth over the past decade due to increased public awareness of privacy and surveillance threats. In response, certain governments are attempting to restrict VPN access by identifying connections using ``dual use'' DPI technology. To investigate the potential for VPN blocking, we develop mechanisms for accurately fingerprinting connections using OpenVPN, the most popular protocol for commercial VPN services. We identify three fingerprints based on protocol features such as byte pattern, packet size, and server response. Playing the role of an attacker who controls the network, we design a two-phase framework that performs passive fingerprinting and active probing in sequence. We evaluate our framework in partnership with a million-user ISP and find that we identify over 85\% of OpenVPN flows with only negligible false positives, suggesting that OpenVPN-based services can be effectively blocked with little collateral damage. Although some commercial VPNs implement countermeasures to avoid detection, our framework successfully identified connections to 34 out of 41 ``obfuscated'' VPN configurations. We discuss the implications of the VPN fingerprintability for different threat models and propose short-term defenses. In the longer term, we urge commercial VPN providers to be more transparent about their obfuscation approaches and to adopt more principled detection countermeasures, such as those developed in censorship circumvention research.

\section{Introduction}
\label{sec:intro}

ISPs, advertisers, and national governments are increasingly disrupting, manipulating, and monitoring Internet traffic~\cite{weaver2011redirecting, li2019large, raman2020investigating, garrett2018monitoring, molavi2015identifying}. As a result, virtual private network (VPN) adoption has been growing rapidly, not only among activists and journalists with heightened threat models but also among average users, who employ VPNs for reasons ranging from protecting their privacy on untrusted networks to circumventing censorship.
As a recent example, with the passage of Hong Kong's new national security law, popular VPN providers observed a 120-fold surge in downloads due to fears of escalating surveillance and censorship~\cite{VPN-HongKong}. 

In response to the growing popularity of VPNs, numerous ISPs and governments are now seeking to track or block VPN traffic in order to maintain visibility and control over the traffic within their jurisdictions. Binxing Fang, the designer of the Great Firewall of China (GFW) said there is an ``eternal war'' between the Firewall and VPNs, and the country has ordered ISPs to report and block personal VPN usage~\cite{vpn-gfw-fang, VPN-China}. More recently, Russia and India have proposed to block VPN services in their countries, both labeling VPNs a national cybersecurity threat~\cite{RKN-VPN, vpn-ban-india}. Commercial ISPs are also motivated to track VPN connections. For example, in early 2021, a large ISP in South Africa, Rain, Ltd., started throttling VPN connections by over 90 percent in order to enforce quality-of-service restrictions in their data plans~\cite{VPN-Rain}. 

ISPs and censors are known to employ a variety of simple anti-VPN techniques, such as tracking connections based on IP reputation, blocking VPN provider (provider from hereon) websites, and enacting laws or terms of service forbidding VPN usage~\cite{seltzer2011infrastructures, tor-in-china, VPN-China}. Yet, these methods are not robust; motivated users find ways to access VPN services in spite of them. However, even less-powerful ISPs and censors now have access to technologies such as carrier-grade deep packet inspection (DPI) with which they can implement more sophisticated modes of detection based on protocol semantics~\cite{filtermap, russia-decentralized}.

In this paper, we explore the implications of DPI for VPN detection and blocking by studying the fingerprintability of OpenVPN (the most popular protocol for commercial VPN services~\cite{Mastering-OpenVPN-Book}) from the perspective of an adversarial ISP\@. We seek to answer two research questions: \emph{(1) can ISPs and governments identify traffic flows as OpenVPN connections in real time}? and \emph{(2) can they do so at-scale without incurring significant collateral damage from false positives?} Answering these questions requires more than just identifying fingerprinting vulnerabilities; although challenging, we need to demonstrate practical exploits under the constraints of how ISPs and nation-state censors operate in the real world.

We build a detection framework that is inspired by the architecture of the Great Firewall~\cite{how-china-detect-shadowsocks, how-china-detect-tor, Ensafi-active-probing}, consisting of \textit{Filter} and \textit{Prober} components. A \textit{Filter} performs passive filtering over passing network traffic in real time, exploiting protocol quirks we identified in OpenVPN's handshake stage. After a flow is flagged by a \textit{Filter}, the destination address is passed to a \textit{Prober} that performs active probing as confirmation. By sending probes carefully designed to elicit protocol-specific behaviors, the \textit{Prober} is able to identify an OpenVPN server using side channels even if the server enables OpenVPN's optional defense against active probing. Our two-phase framework is capable of processing ISP-scale traffic at line-speed with an extremely low false positive rate.

In addition to core or ``vanilla'' OpenVPN, we also include commercial ``obfuscated'' VPN services in this study. In response to increasing interference from ISPs and censors, obfuscated VPN services have started to gain traction, especially from users in countries with heavy censorship or laws against the personal usage of VPNs. Obfuscated VPN services, whose operators often tout them as ``invisible'' and ``unblockable''~\cite{TorGuard-Obfuscated, Surfshark-Obfuscated, Boleh-Obfuscated}, typically use OpenVPN with an additional obfuscation layer to avoid detection~\cite{Vypr-Obfuscated, Astrill-obfs}.

\begin{figure}[t]
\centering
 \includegraphics[width=\columnwidth,keepaspectratio]{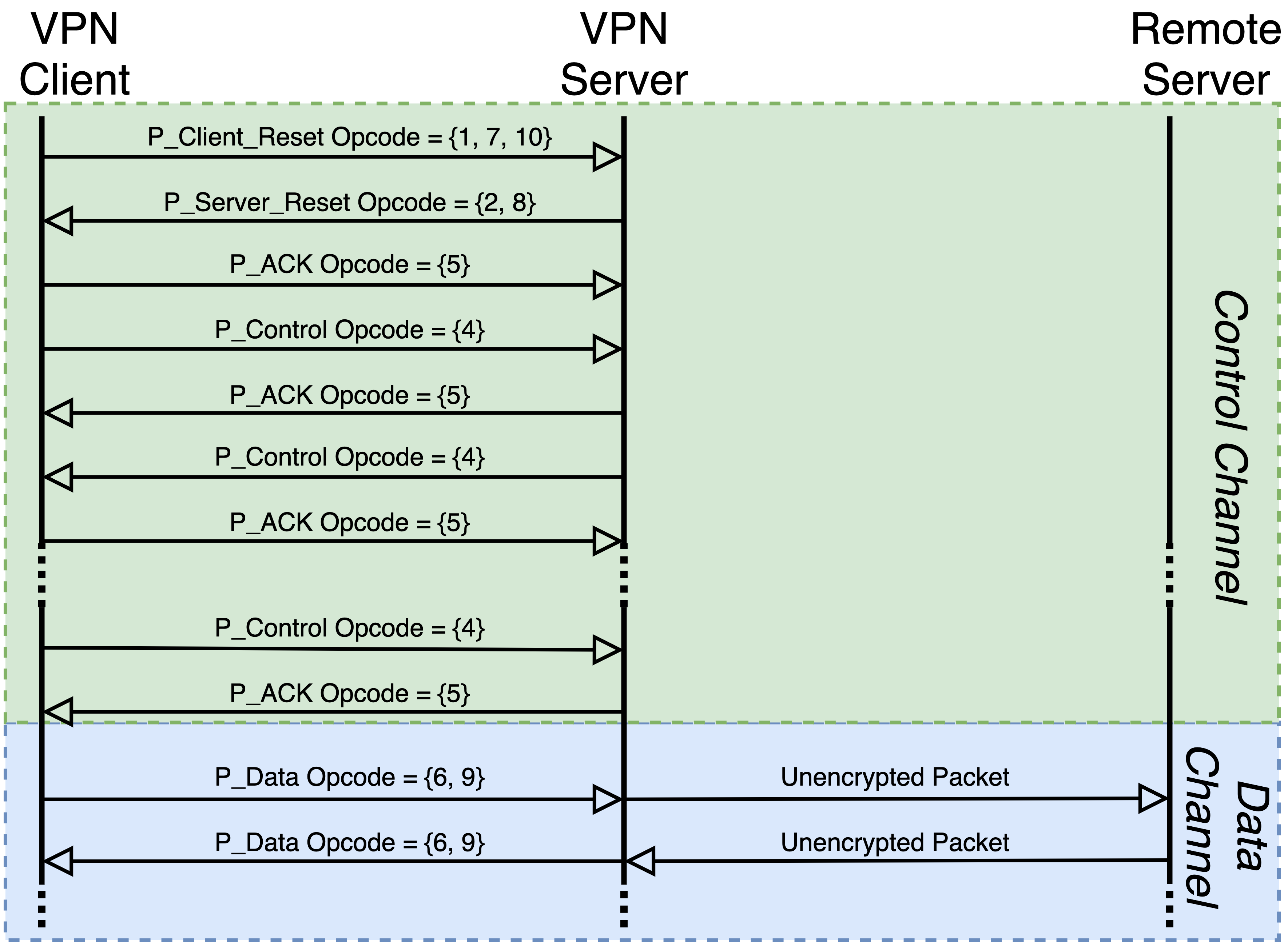}
\caption{\textbf{OpenVPN Session Establishment (TLS mode).}}
\label{fig:negotiation}
\vspace{-15pt}
\end{figure}


Partnering with \textit{Merit} (a mid-size regional ISP that serves a population of 1 million users), we deploy our framework at a monitor server that observes 20\,Gbps of ingress and egress traffic mirrored from a major \textit{Merit} point-of-presence. (Refer to \S~\ref{sec:ethics} for ethical considerations.) We use PF\_RING~\cite{pf-ring-zc} in zero-copy mode for fast packet processing by parallelized \textit{Filters}. In our tests, we are able to identify 1718 out of 2000 flows originating from a control client machine residing within the network, corresponding to 39 out of 40 unique ``vanilla'' OpenVPN configurations.

More strikingly, we also successfully identify over two-thirds of obfuscated OpenVPN flows. Eight out of the top 10 providers offer obfuscated services, yet \textit{all} of them are flagged by our \textit{Filter}. Despite providers' lofty unobservability claims (such as ``$\ldots$ even your Internet provider can’t tell that you’re using a VPN''~\cite{Surfshark-Obfuscated}), we find most implementations of obfuscated services resemble OpenVPN masked with the simple XOR-Patch~\cite{OpenVPN-XOR}, which is easily fingerprintable. Lack of random padding at the obfuscation layer and co-location with vanilla OpenVPN servers also make the obfuscated services more vulnerable to detection.

In a typical day, our single-server setup analyzes 15\,TB of traffic and 2~billion flows. Over an eight-day evaluation, our framework flagged  3,638 flows as OpenVPN connections. Among these, we are able to find evidence that supports our detection results for 3,245 flows, suggesting an upper-bound false-positive rate three orders of magnitude lower than previous ML-based approaches~\cite{perceptron-nn-VPN, VPN-Comparison, LSTM-Attention}.
\looseness=-1

We conclude that tracking and blocking the use of OpenVPN, even with most current obfuscation methods, is straightforward and within the reach of any ISP or network operator, as well as nation-state adversaries. Unlike circumvention tools such as Tor or Refraction Networking~\cite{Tor-original-paper, Wustrow14-Tapdance}, which employ sophisticated strategies to avoid detection, robust obfuscation techniques have been conspicuously absent from OpenVPN and the broader VPN ecosystem. For average users, this means that they may face blocking or throttling from ISPs, but for high-profile, sensitive users, this fingerprintability may lead to follow-up attacks that aim to compromise the security of OpenVPN tunnels~\cite{Jed-VPN-Attack, port-shadowing}. We warn users with heightened threat models \emph{not} to expect that their VPN usage will be unobservable, even when connected to obfuscated services. While we propose several short-term defenses for the fingerprinting exploits described in this paper, we fear that, in the long term, a cat-and-mouse game similar to the one between the Great Firewall and Tor is imminent in the VPN ecosystem as well. We implore VPN developers and providers to develop, standardize, and adopt robust, well-validated obfuscation strategies and to adapt them as the threats posed by adversaries continue to evolve.

\section{Background \& Related Work}
\label{sec:background}

VPN tools create private networks across the public Internet through encrypted tunneling. Although many VPN protocols are being used, such as IPSec and WireGuard, OpenVPN remains the most supported and trusted protocol among commercial VPN providers~\cite{Mastering-OpenVPN-Book}. Due to its versatility and open-source nature, OpenVPN has been used as the underlying protocol in numerous VPN products, which often advertise the protocol for its proven security~\cite{Vypr-Obfuscated}. In addition, OpenVPN's popularity continues to rise with the trend of users choosing to self-host open-source VPN tools~\cite{VPN-Trend}.

\paragraph{OpenVPN Protocol.} OpenVPN was first released in 2002 with the aim of creating a tunneling protocol focusing on security, while also being free and fast over the standard TCP and UDP~\cite{OpenVPN-History}. When the OpenVPN tunnel is active, raw IP packets being sent to or from the tunnel to the final destination are encapsulated inside OpenVPN packets. To achieve secure communication, OpenVPN leverages the OpenSSL library as its cryptographic layer. Two methods for authentication and key exchange are provided to establish trust with peers: either pre-shared static key(s) or TLS-based negotiations. The latter has been adopted by the majority of commercial VPN services.
Two separate channels are used for key exchange and data transfer, both sharing a single multiplexed TCP/UDP stream. In the control channel, the client and server engage in a TLS-style exchange of key materials. As TLS is designed to operate over a reliable transport, OpenVPN provides its control channel with a sequential, reliable layer based on an explicit acknowledgement and re-transmission mechanism. The negotiated key from the control channel will be used to encrypt packets transferred in the data channel, which does not provide any reliability guarantee. Figure~\ref{fig:negotiation} presents a typical initialization sequence of OpenVPN packets leading to a fully encrypted data channel.

\paragraph{Tor, Proxy, and VPN Detection.} The ongoing arms-race between the GFW and Tor has been extensively studied and is most representative of the conflict between censorship \& surveillance and circumvention tools~\cite{Ensafi-active-probing,how-china-detect-tor, sok-obfs, Ensafi-GFW-Space-Time, Gill-Foci18-Tor, Tor-tls-block-blog}. Censors started by blocking Tor's website and public relays, which Tor responded to by deploying website mirrors and private, unpublished bridges. Next, censors moved to blocking with DPI by fingerprinting Tor's TLS handshake, \eg cipher suites. Tor used Pluggable Transports (PT) obfuscators, such as Obfsproxy and meek~\cite{Pluggable-Transports-Project}, to mask the handshake. In response, censors deployed active probing to complement DPI-based fingerprinting to detect Tor and certain obfuscators.
\looseness=-1

There is limited previous work focusing on VPN traffic detection. Hoogstraaten~\cite{hoogstraaten2018evaluating} explored server-side VPN detection methods, ranging from using existing information databases (\eg WHOIS, rDNS) to fingerprinting TCP options (\eg advertised MSS). Webb \etal~\cite{webb2016finding} proposed detecting proxies and VPNs based on traffic timing and latency. Their approach relied on the hypothesis that when a service is accessed through a proxy, the RTT measurement will be different from the RTT of a direct connection. Another class of previous work uses computational and machine learning models to passively detect VPN traffic~\cite{VPN-nonVPN, 1d-cnn-vpn, Deep-Packets, perceptron-nn-VPN, VPN-Comparison, LSTM-Attention, payload-length-sequence}, leveraging flow-level statistics such as connection duration and packet interval. Most of this work uses the same synthetic \texttt{ISCXVPN2016} dataset~\cite{VPN-nonVPN}---which contains a balanced mixture of VPN and non-VPN traffic---to train and test a variety of machine learning and neural network classifiers in an offline, lab-setting. In contrast, our work primarily focuses on whether ISP-level adversaries can identify OpenVPN flows in near real time, and whether they can do so at scale, under practical constraints, and with minimal collateral damage. For this reason, we omit a full analysis of ML-based work, and only compare them with our approach in terms of false positives (falsely blocking legitimate traffic).


\paragraph{Obfuscated (Open)VPN.} Various traffic obfuscation techniques have been examined in previous work. Wang~\etal examined the detectability of Obfsproxy, FTE, and meek~\cite{detect-obfuscation}. Using attacks based on protocol semantics, packet entropy, and timing-related features, they concluded that a determined censor could detect all three obfuscators reliably. Houmansadr~\etal demonstrated that popular mimicry-based obfuscation tools failed to achieve unobservability because seamlessly simulating another protocol is extremely challenging~\cite{Houmansadr2013}. Previous studies have suggested censors can use active probing to detect proxies that obfuscate traffic~\cite{Ensafi-active-probing, how-china-detect-shadowsocks, how-china-detect-tor}. In response, ``probe-resistant'' proxies were developed, which remain silent when being probed by an unauthenticated adversary. However, researchers have demonstrated that carefully designed probes could still identify these proxies~\cite{Frolov-probe-resistant}.

There is a marked demand for an emerging class of services called ``stealth'' or ``obfuscated'' VPN, especially from users in countries with heavy censorship or laws against personal VPN usage~\cite{VPN-Pakistan, VPN-China}. Most obfuscated VPN services use OpenVPN as the underlying protocol for security and routing, with an obfuscation layer overlaid to avoid detection~\cite{Vypr-Obfuscated, Astrill-obfs}~\footnote{There are discussions on obfuscating WireGuard~\cite{wireguard-obfuscation-1, wireguard-obfuscation-2}, but to the best of our knowledge, they have yet to be deployed by any commercial VPNs}. OpenVPN's core developers prefer that obfuscation remains a separate project operating alongside the vanilla/core protocol, as they ``do not want to play the cat-and-mouse game [as Tor]''~\cite{obfuscation-thread}. The absence of a standardized obfuscation solution has led to a plethora of obfuscators implemented by different VPN providers, who often claim that their obfuscated services can remain undetected by ISPs and censors alike. For example, TorGuard introduces their obfuscated VPN service as ``Engineered from the ground up to be impossible to detect''~\cite{TorGuard-Obfuscated}. BolehVPN claims that their VPN obfuscation ``$\ldots$keeps you out of trouble, even in China''~\cite{Boleh-Obfuscated}. Common obfuscation strategies adopted by commercial VPNs include employing XOR-based scramblers, wrapping OpenVPN inside encrypted tunnels, or using proprietary protocols.
\looseness=-1

\textbf{OpenVPN XOR Patch:} Originally developed by Clayface as a patch for vanilla OpenVPN, the XOR patch scrambles a packet by either xor-ing bytes with a pre-shared key, reversing the order of the bytes, xor-ing each byte with its position, or a combination of these steps~\cite{OpenVPN-XOR}. Notably, OpenVPN developers discourage its use due to the lack of code audit~\cite{Tunnelblick-XOR}. 

\textbf{OpenVPN over Encrypted Tunnels:} Some VPN services wrap OpenVPN traffic inside encrypted tunnels to prevent DPI fingerprinting. Some of the adopted obfuscation tunnels are Obfsproxy (obfs\{2/3/4\}), Stunnel, Websocket Tunnel, and encrypted proxies (shadowsocks, V2Ray).

\textbf{Proprietary Protocols:} A few VPN providers have developed proprietary obfuscated protocols, some of which are built on top of OpenVPN with a proprietary obfuscation layer added, such as VyprVPN or Astrill~\cite{Vypr-Obfuscated, Astrill-obfs}.

To the best of our knowledge, we are the first to explore the fingerprintability of commercial and/or obfuscated OpenVPN services on real traffic. Our unique study highlights the practicality of such fingerprinting, which has profound real-world security implications on end-users expecting certain privacy and anonymity guarantees from using these services.


\begin{figure*}[t]
\centering
 \includegraphics[width=2\columnwidth,keepaspectratio]{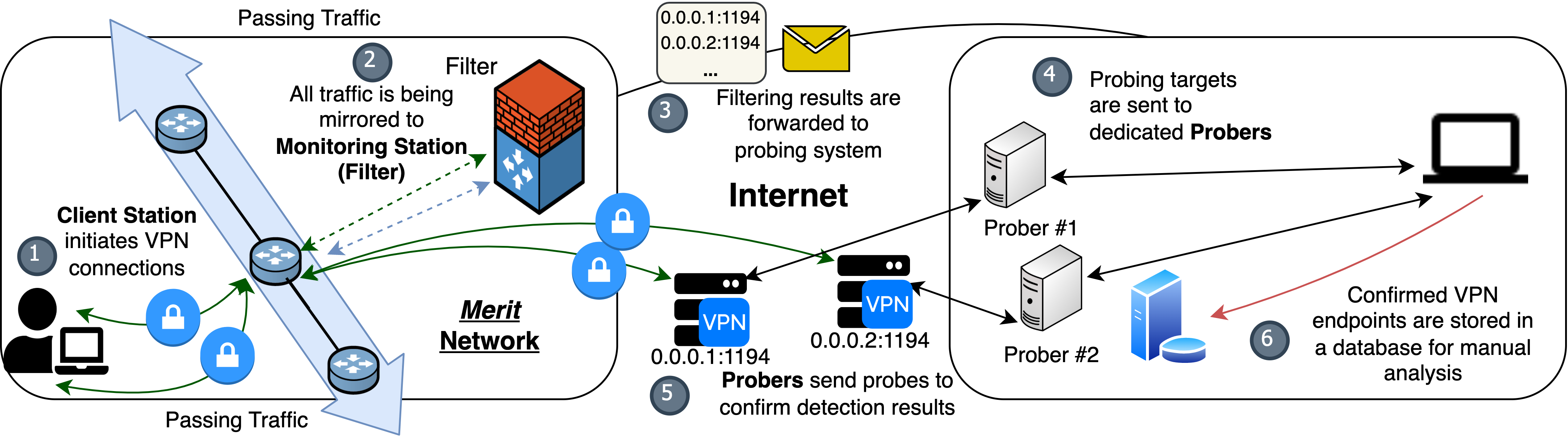}
\caption{\textbf{Framework Deployment on \textit{Merit}} Steps: (1) Client connects to VPN servers. (2) VPN connections, along with passing traffic, are being mirrored to the \textit{Filter}. (3) \textit{Filter} forwards server IP of suspected connections to the probing system. (4) Targets are sent to each dedicated \textit{Probers}. (5) \textit{Probers} send probes asynchronously. (6) Connections confirmed by probing are logged.}
\label{fig:deployment}
\vspace{-5pt}
\end{figure*}
\section{Challenges in Real-world VPN Detection}
\textbf{Effective investigation of fingerprintability requires incorporating perspectives of how ISPs and censors operate in practice.} It is not enough to simply identify fingerprinting vulnerabilities, we need to demonstrate realistic exploits to illustrate the practicality of exploiting the vulnerability, while taking into consideration the ISP and censors' capabilities and constraints~\cite{sok-obfs}. For instance, previous academic works considered using flow-level features to train ML classifiers for VPN detection~\cite{VPN-nonVPN, 1d-cnn-vpn, Deep-Packets, perceptron-nn-VPN, VPN-Comparison, LSTM-Attention}. Yet, it remains unclear how practical these detection approaches are for ISPs and censors, and we know of no rigorous studies that examine real-world deployment of an ML-based censorship system~\cite{sok-obfs}. Furthermore, previous works test on the \texttt{ISCXVPN2016} dataset~\cite{VPN-nonVPN} with balanced OpenVPN and non-VPN traffic. However, we note that due to the low base rate of VPN traffic in the wild, even the best-performing ML system has false positive rates that can be economically impractical for real-world censors sensitive to collateral damage~\cite{detect-obfuscation}. 

However, investigations adopting the viewpoint of ISPs and censors can be challenging. First, such investigation requires collaboration with real-world ISPs and access to their network traffic. We need to install monitors inside an ISP's network, while ensuring our analysis will not affect ISP's normal routing operations. Furthermore, analyzing traffic from real users raises ethical concerns. Processing raw network data may violate the privacy of users, in particular VPN users who often have a heightened threat model. Finally, deploying a system that performs ad-hoc traffic analysis in real time poses significant engineering challenges. We need to ensure the entire analysis framework (including processing and logging) keeps pace with the packet arrival rate and take into consideration the effect of potential asymmetric routing or packet loss on the analysis and results.  

\section{Adversary Model and Deployment}

We assume a realistic censor (ISP) capability model based on knowledge from previous measurement studies on the arms race between censors and circumventors~\cite{how-china-detect-shadowsocks, how-china-detect-tor, sok-obfs, Ensafi-active-probing}. We outline a censor-controlled on-path filter that passively observes and examines passing network traffic. The filter is stateful, but has limited resources and can maintain a limited amount of per-connection states for a short time. The filter is also constrained by long-term data storage and computational resources. In addition to filters installed inside the monitored networks, we assume the censor also operates measurement machines that can send protocol-specific probes to further confirm the detection result. Such two-phase systems have already been adopted by real-world censors such as the GFW against Tor and Shadowsocks~\cite{how-china-detect-tor, how-china-detect-shadowsocks}. Finally, we expect the censor is familiar with the protocol of interest and has access to the different obfuscators deployed by VPN providers (\eg, as a paid customer). We emphasize that this threat model corresponds to censor's capabilities as observed in practice \textit{today}, rather than future capabilities.


To investigate the fingerprintability of OpenVPN and existing obfuscated solutions, we set up a two-phase detection framework in order to answer our key questions: 1) whether real-world censors are \textit{capable} of performing such detection, and 2) whether it is economical to do this \textit{at scale}. Figure~\ref{fig:deployment} shows an overview of our framework deployment. Partnering with \textit{Merit}, we instantiate a \textit{Filter} on a Monitoring Station overseeing mirrored traffic from a router that handles 20\% of the ISP's traffic. 
The \textit{Filter} performs passive fingerprinting over raw packets, exploiting traffic features unique to OpenVPN. IP and port information of flows flagged by the \textit{Filter} are forwarded to a probing system and then distributed to dedicated \textit{Probers}. The \textit{Probers} send a set of pre-defined probes specifically designed to fingerprint an OpenVPN server. Finally, probed servers that are confirmed as OpenVPN are logged for manual analysis. Such a two-phase framework resembles how real-world censors operate: lightweight filtering followed up by more expensive, but also more accurate, active probing. This framework is capable of processing massive traffic in real-time while also preventing excessive collateral damage.
\looseness=-1

\section{Ethics, Privacy, and Responsible Disclosure}
\label{sec:ethics}

Raw network traffic that contains real users' data is highly sensitive, and this is especially true for traffic related to privacy-oriented services such as VPNs. Here we describe how we consider the security and privacy risks and ethical issues raised by our work, and we detail the procedural and technical steps we take to mitigate the risks.

Foremost among the ethical concerns associated with this work is our \textit{Filter} deployment inside \textit{Merit}'s network to analyze user traffic. \textit{Merit}, which has extensive previous experience collaborating with universities and has well-defined ethics and privacy rules to govern such projects, supervised the deployment. We also cleared our research plan with our university legal counsel and IRB\@. Although the IRB determined that the work is not regulated, we take extensive measures to minimize potential risks for end-users.

Our framework is fine-tuned on both real and lab-generated traffic data, and it is evaluated on live ISP traffic. For controlled fine-tuning, a small traffic snapshot (the ISP Dataset in section~\ref{sec:finetune}) was used to calibrate parameters, \eg, the size of observation window. The traffic snapshot, sampling 1/30 of all flows for 45 minutes on July 28, 2021, was generated and analyzed entirely on \textit{Merit} systems, with security mechanisms limiting access to select members of the team. As with the design described in Section~\ref{sec:Filter}, \textit{Filter} analyzed only the first payload byte, completely ignoring the remainder of the payload, and it recorded only the observed degree of variation. The raw snapshot was never inspected by humans and was deleted after the fine-tuning concluded.

For deployment and evaluation on live ISP traffic, the \textit{Filter} architecture is designed to minimize risks of disrupting or modifying user traffic. The Monitoring Station only receives a copy of the traffic, so even if our software were to malfunction, network service would be unaffected.
In addition, to reduce privacy risks, the \textit{Filter} collects only the minimum information necessary for the subsequent probing operation. It records only the server IP addresses and ports of matching connections, which are bucketed into 5-minute internals to inhibit time correlation. These logs are stored and analyzed on a server that is securely maintained by \textit{Merit} and is accessible only to a few members of our research team on a least-privilege basis. \textit{Merit} reviewed our source code prior to deploying it on their network. During deployment and evaluation, no packet payloads or client IP addresses are ever recorded to disk or inspected by humans.

\begin{figure}[t]
\centering
 \includegraphics[width=\columnwidth,keepaspectratio]{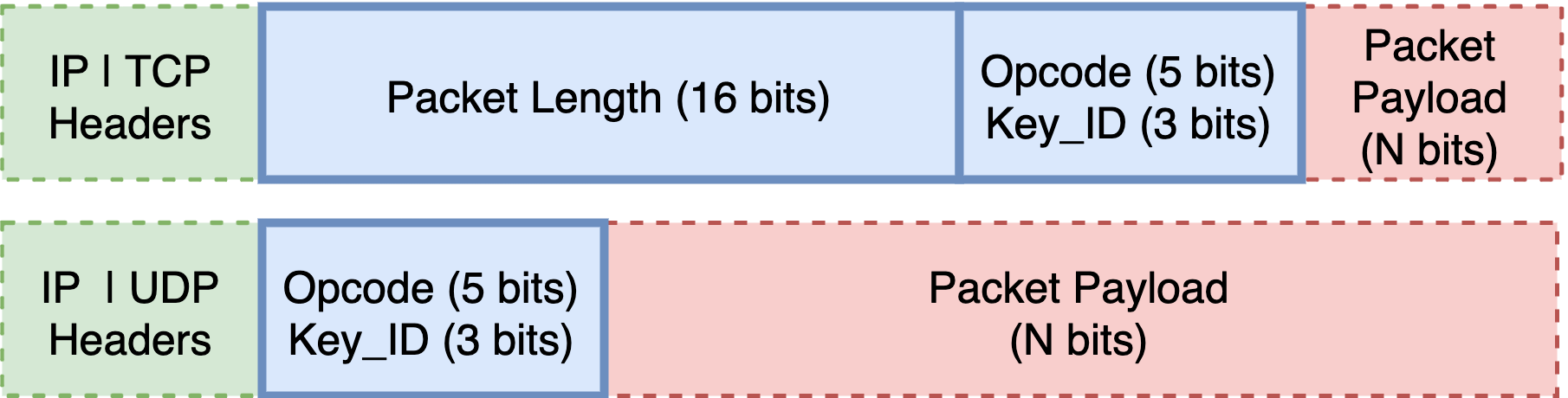}
\caption{\textbf{OpenVPN Header in TCP and UDP modes.} (TLS only)}
\label{fig:openvpn-header}
\vspace{-15pt}
\end{figure}

Based on the \textit{Filter} log, the \textit{Probers} send probes to candidate VPN servers. To minimize the risk of disrupting server operations, we design the probes to be non-invasive and make information available to assist operators in debugging any problems we inadvertently cause. Each server receives only 2--10 innocuous connection attempts, similar to those commonly used in Internet measurement tools like Nmap. The probes originate from two dedicated machines that we provisioned with web pages that explain the nature of the experiment and provide our contact information. We did not receive any inquiries, complaints, or problem reports.
Since the server IP addresses themselves may sometimes be non-public, we only report aggregate statistics (\eg, the false positive rate) and will not publish any of the addresses that we collect. Any data requests will be referred to \textit{Merit}.

As with all attack-oriented research, there is a risk that our work developing VPN fingerprinting techniques will be adopted by real attackers. To minimize this risk, we are in the process of responsibly disclosing our findings to the VPN operators whose obfuscated servers we successfully identified in our evaluation. We believe that the security of the VPN ecosystem is best advanced by having these problems surfaced by responsible researchers. Our work will help accurately inform users about the VPN services they rely on, and we hope it will enable more robust countermeasures to be developed and deployed.



\section{Identifying Fingerprintable Features}
\label{sec:Filter}
In this section, we identify three features that fingerprint OpenVPN, exploiting byte pattern, packet length, and server behaviors, respectively.

\subsection{Opcode-based Fingerprinting}

As shown in Figure~\ref{fig:openvpn-header}, each OpenVPN packet has a header of 24 bits in TCP mode or 8 bits in UDP mode, which is not part of the encrypted payload. Each OpenVPN header starts with an opcode that specifies the message type of the current packet and a key ID that refers to a (new) TLS session. The opcode field can take over 10 defined values, corresponding to message types transmitted during different communication stages. A typical OpenVPN session starts with the client sending a \texttt{Client Reset} packet. The server then responds with a \texttt{Server Reset} packet, and a TLS handshake follows. OpenVPN packets that carry TLS ciphertexts have \texttt{P_Control} as their message type. Since OpenVPN can run over UDP but has to provide a reliable channel for TLS, each \texttt{P_Control} packet is explicitly acknowledged by \texttt{P_ACK} packets. Finally, actual payloads are transmitted as \texttt{P_Data} packets. Figure~\ref{fig:negotiation} illustrates this packet exchange with opcode annotations.

\begin{algorithm}[t]
\label{alg:filter-algo}
\caption{Opcode Fingerprinting Logic}
\begin{algorithmic}
\Require $N \geq 0$
\State $OCSet \gets \{\}$, $CR \gets Opcode[0]$, $SR \gets Opcode[1]$
\State $i \gets 2$
\While{$i \neq N \And i < |Opcode|$}
\If{$Opcode[i]$ $\in$ {$CR$,$SR$} $\And$ |$OCSet$| $\geq$ 4}
\State \textbf{Return} False
\EndIf
\State $OCSet$ += $Opcode[i]$ 
\State $i \gets i + 1$
\EndWhile
\State \textbf{Return} $i == N \And 4 \leq$ |$OCSet$| $\leq$ 10
\State \#At least 4 different Opcodes needed to complete handshake. In total 10 Opcodes defined by the protocol.
\end{algorithmic}
\label{alg:opcode-fingerprinting}
\end{algorithm}
A packet field taking a fixed number of values can be easy to fingerprint and has been exploited before against other protocols~\cite{how-china-detect-shadowsocks}. We fingerprint OpenVPN's handshake sequence by analyzing each opcode byte for the first \textit{N} packets of a flow (the threshold \textit{N} is explored in Section~\ref{sec:Observation-Window}). Algorithm~\ref{alg:opcode-fingerprinting} shows the process of opcode fingerprinting, with \texttt{Opcode} referring to the sequence of \textit{N} opcode values found in the first \textit{N} packets of a given flow. Briefly, the filter flags a flow if the number of different opcodes observed accords with the protocol \textit{and} the \texttt{Client} and \texttt{Server Resets} are not seen once the handshake is completed.

\begin{figure}[t]
\centering
 \includegraphics[width=\columnwidth,keepaspectratio]{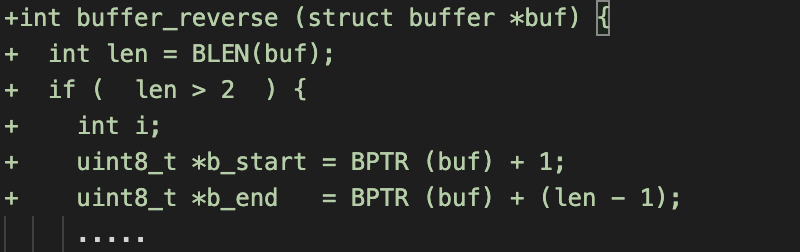}
\caption{\textbf{XOR-Patch that leaves first byte un-reversed}}
\label{fig:xor-reverse}
\vspace{-15pt}
\end{figure}

Previous work and existing open-source DPIs~\cite{ndpi, Zeek, libprotoident, OpenVPN-Opcode} considered statically matching opcode values and packet sizes based on the protocol specification. In contrast, we propose to dynamically capture the variation in opcode values that reflects the establishment of OpenVPN sessions. Notably, our heuristics do not require exact matching of opcode values or packet length (\eg, do not require the third byte of the first packet to be 0x38), thereby ensuring it works effectively against XOR-obfuscated flows. The XOR obfuscation masks packet payloads to ensure that the opcode bytes are altered. Notably, according to the specification~\cite{OpenVPN-XOR}, when it reverses the packet as one of the obfuscation steps, it excludes the first character of the buffer (where the opcode byte is located) from reversal, as shown in Figure~\ref{fig:xor-reverse}. As such, the opcode byte is always XOR-ed with the same byte of the XOR key, and the same opcodes would be mapped to the same value after obfuscation. This behavior is preserved when Tunnelblick (a popular OpenVPN client on macOS) adopts the patch~\cite{Tunnelblick-XOR}, and has been used in multiple mobile apps~\cite{Zhang2017OhPwnVPNSA}. By considering only the number of unique opcodes seen so far, our heuristics are more flexible and target various XOR-based obfuscations of OpenVPN.

\subsection{ACK-based Fingerprinting}

OpenVPN engages in a TLS-style handshake with its peer over the control channel. Since TLS is designed to operate over a reliable layer, OpenVPN implements an explicit acknowledgement and re-transmission mechanism for its control channel messages~\cite{openvpn-reliable}. Specifically, incoming \texttt{P_Control} packets are acknowledged by \texttt{P_ACK} packets, which do not carry any TLS payloads and are uniform in size (Note these ACK packets are carried over by TCP as payload and are not the same as TCP ACK flags). Moreover, these ACK packets are seen mostly only in the early stage of a flow, during the handshake phase, and are not used in the actual data transfer channel, which can run over an unreliable layer.

To our knowledge, we are the first to devise fingerprinting attempts based on the distinct protocol-layer ACKs against OpenVPN. Previously, the unique timing pattern in meek's TCP-level ACK traffic has rendered the obfuscation tool vulnerable to detection~\cite{detect-obfuscation}. For OpenVPN, the presence of explicit ACK packets, uniform in size and only seen in some parts of a session, provides another fingerprintable feature. Specifically, we first identify a likely ACK packet of a session by locating an initial packet exchange sequence of C->S (Client-Reset), S->C (Server-Reset), C->S (ACK), C->S (Control), as illustrated in Figure~\ref{fig:negotiation}. For vanilla OpenVPN and XOR-based obfuscation, the first ACK packet usually appears as the third (data) packet transmitted in a session. For tunnels or obfuscators that have their own handshake or key exchange process (\eg, Stunnel, SSH tunnel, or Obfsproxy), this counting is offset by the number of tunnel handshake packets. Next, we group packets into 10-packet bins, and we derive the ACK fingerprint for each flow by counting the number of packets in each bin that have the same size as the identified ACK packet. For OpenVPN flows, we expect to observe a high number of ACK packets in early bins and an absence of them in later bins. (Later in the session, Control and ACK packets can be exchanged again to transfer random key materials, but it is not expected to be observed within our observation window N.) This approach proves effective to fingerprint vanilla OpenVPN as well as obfuscated services running over encrypted tunnels that lack random padding. We quantify exact fingerprinting thresholds in Section~\ref{ACK-Threshold}.

\subsection{Active Server Fingerprinting}
\label{active-probing}

We explore the feasibility of identifying an OpenVPN server through active probing. Typically, OpenVPN servers respond to a client reset with an explicit server reset, thereby giving away their identity. However, most commercial providers now have adopted \textit{tls-auth} or \textit{tls-crypt} options~\cite{tls-auth}. These options add an additional HMAC signature---signed by a pre-shared key---to every control channel packet for integrity verification, including the initial reset packets. With either of these options enabled, an OpenVPN server would not respond to an unauthenticated client reset with a server reset, but would instead drop such packets without further processing. 
The presence of such HMAC mechanism increases the complexity of doing active probing: it effectively makes OpenVPN servers ``probe-resistant''~\cite{Frolov-probe-resistant} by remaining silent when probed by an unauthenticated client.

In fact, similar HMAC mechanisms are used by more popular ``probe-resistant'' proxies, such as obfs4~\cite{obfs4}. However, unlike obfs4 which waits for a server-specific random delay before dropping an unauthenticated connection, OpenVPN always \textit{immediately} closes the connection if a valid HMAC cannot be located. We design our probes to leverage this protocol-specific behavior, and as a result, we manage to fingerprint OpenVPN servers even if they do not respond throughout our probing cycles. The key concept is that although the application may not respond to probing, an attacker may still be able to fingerprint application-specific thresholds at the TCP level, such as timeouts or RST thresholds, as demonstrated by Frolov \etal~\cite{Frolov-probe-resistant}.

We use two datasets in this section to help with designing probes. \textbf{ZMap Set}: to construct a realistic non-VPN endpoints dataset, we use ZMap to scan each of the 65,535 TCP ports over the entire IPv4 space, limiting results for each port to 200 endpoints (with the specific port open), resulting in over 13 million endpoints. \textbf{Censys Set}: We query the \texttt{Censys.io}~\cite{censys15} database for hosts with TCP port 1194/OpenVPN open. Next, we probe each endpoint with a typical OpenVPN \texttt{Client Reset} and group endpoints that respond with explicit \texttt{Server Resets}. This results in 180,858 hosts known to be OpenVPN endpoints (with ``tls-auth'' disabled).
\subsubsection{Base Probes}
\begin{figure}[t]
\centering
 \includegraphics[width=\columnwidth,keepaspectratio]{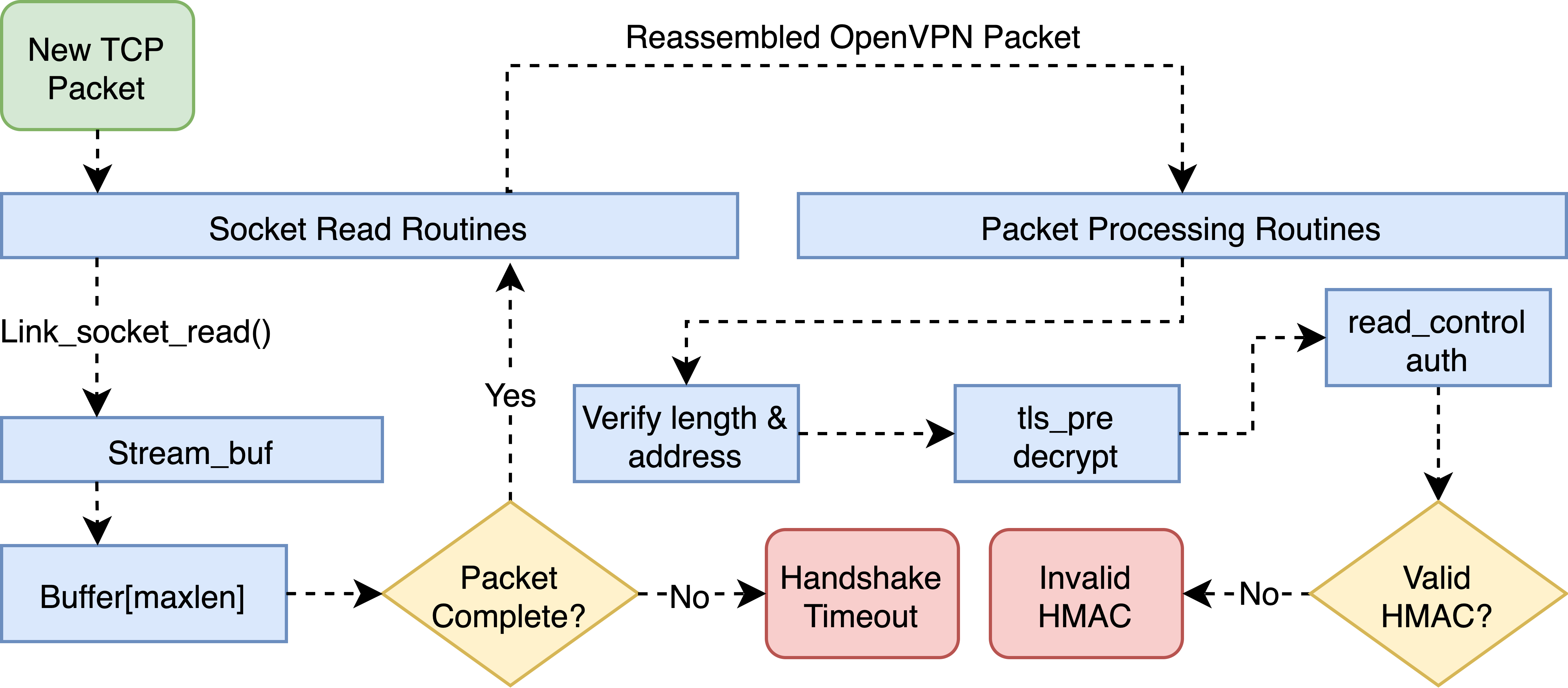}
\caption{\textbf{OpenVPN TCP new packet processing routines}}
\label{fig:openvpn-tcp}
\vspace{-5pt}
\end{figure}
\begin{table}[]
\footnotesize
\begin{tabular}{p{0.17\columnwidth}p{0.45\columnwidth}p{0.25\columnwidth}}
\toprule
ProbeName     & Probe Content                             & Expected Behavior            \\
\midrule
BaseProbe 1 & x00x0ex38.\{8\}x00x00x00x00x00      & Explicit ServerReset or 
Short Close  \\
BaseProbe 2 & x00x0ex38.\{8\}x00x00x00x00          & Long Close 
\\ 
TCP Generic   & x0dx0ax0dx0a                          & Short Close 
\\ 
One Zero      & x00                                      & Long Close 
\\ 
Two Zero      & x00x00                                  & Short Close 
\\ 
Epmd          & x00x01x6e                              & Short Close 
\\ 
SSH           & SSH-2.0-OpenSSH_8.1/r/n                   & Short Close 
\\ 
HTTP-GET      & GET/HTTP/1.0 /r /n /r /n                    & Short Close 
\\ 
TLS           & Typical Client Hello by Chromium & Short Close 
\\ 
2K-Random           & Random 2000 Bytes & Short Close \& RST 
\\ 
\bottomrule
\end{tabular}
\caption{\textbf{Summary of Probes and the expected behaviors from an OpenVPN server.}}
\label{tab:probes}
\vspace{-15pt}
\end{table}

We design probes exploiting a behavior associated with how OpenVPN packetizes TCP streams. When OpenVPN operates over TCP, it needs to split the continuous stream into discrete OpenVPN packets. Figure~\ref{fig:openvpn-tcp} presents a high-level abstraction of this process. The most relevant parts are: a buffer is allocated in memory to reassemble fragments of OpenVPN packets encapsulated in TCP streams. The length N for the next OpenVPN packet is extracted from the first two bytes of the header (see Figure~\ref{fig:openvpn-header}), and the routine keeps reading N additional bytes before it returns the reassembled packet to the caller. This means that an OpenVPN packet will not be parsed and checked for syntax and encryption errors until all its parts arrive at the server. Based on this behavior, we design two sequential probes to trigger an OpenVPN server into different code paths---which result in different connection timeouts---and measure the time elapsed before the server responds or terminates the connection. As shown in Table~\ref{tab:probes}, \textit{Base Probe 1} carries a typical 16-byte OpenVPN \texttt{Client Reset}, while \textit{Base Probe 2} has the same payload with the last byte stripped off. The assumption is since our two probes only differ in one byte, most non-OpenVPN servers will respond to our probes in a similar way. However, for an OpenVPN server with HMAC enabled, the connection sending the first probe will be dropped immediately because the OpenVPN packet is reassembled and a valid HMAC cannot be located. The second probe will not receive an immediate response, as the server will wait for an additional byte to arrive for reassembly. The connection will stay idle until a server specific handshake timeout has passed, after which the connection will be dropped. As such, the first probe will be dropped at the decryption routine, while the second probe will be dropped at the packet reassembly routine, both labeled red in Figure~\ref{fig:openvpn-tcp}.

\subsubsection{Additional Probes}

The two probes, although useful, are limited and there may be other protocols with behaviors similar to OpenVPN. After using both to probe the \textit{ZMap Set}, we still identify a handful of services that respond similarly to OpenVPN servers, such as Microsoft WBT Server (3389), Microsoft Message Queuing (1801), and Erlang Port Mapper Daemon (4369). 

We design additional probes based on the fact that OpenVPN validates packet length and will drop connections sending invalid length without waiting for the next packet to be reassembled.
Here, packet length refers to the length declared by the first two bytes of an OpenVPN header (see Figure~\ref{fig:openvpn-header}), rather than the TCP packet length. A ``valid'' length is in the range of [1, \textit{max_len}], where \textit{max_len} is derived from the server's MTU configurations. For instance, default TUN MTU of 1500 bytes, combined with overheads (crypto IV, packet length, \etc), results in a \textit{max_len} of 1627 bytes. In this case, probes whose first two bytes have a decimal value greater than 1627 (0x06,0x5B) will be dropped immediately.

\begin{figure}[t]
\centering
 \includegraphics[width=1.03\columnwidth,keepaspectratio]{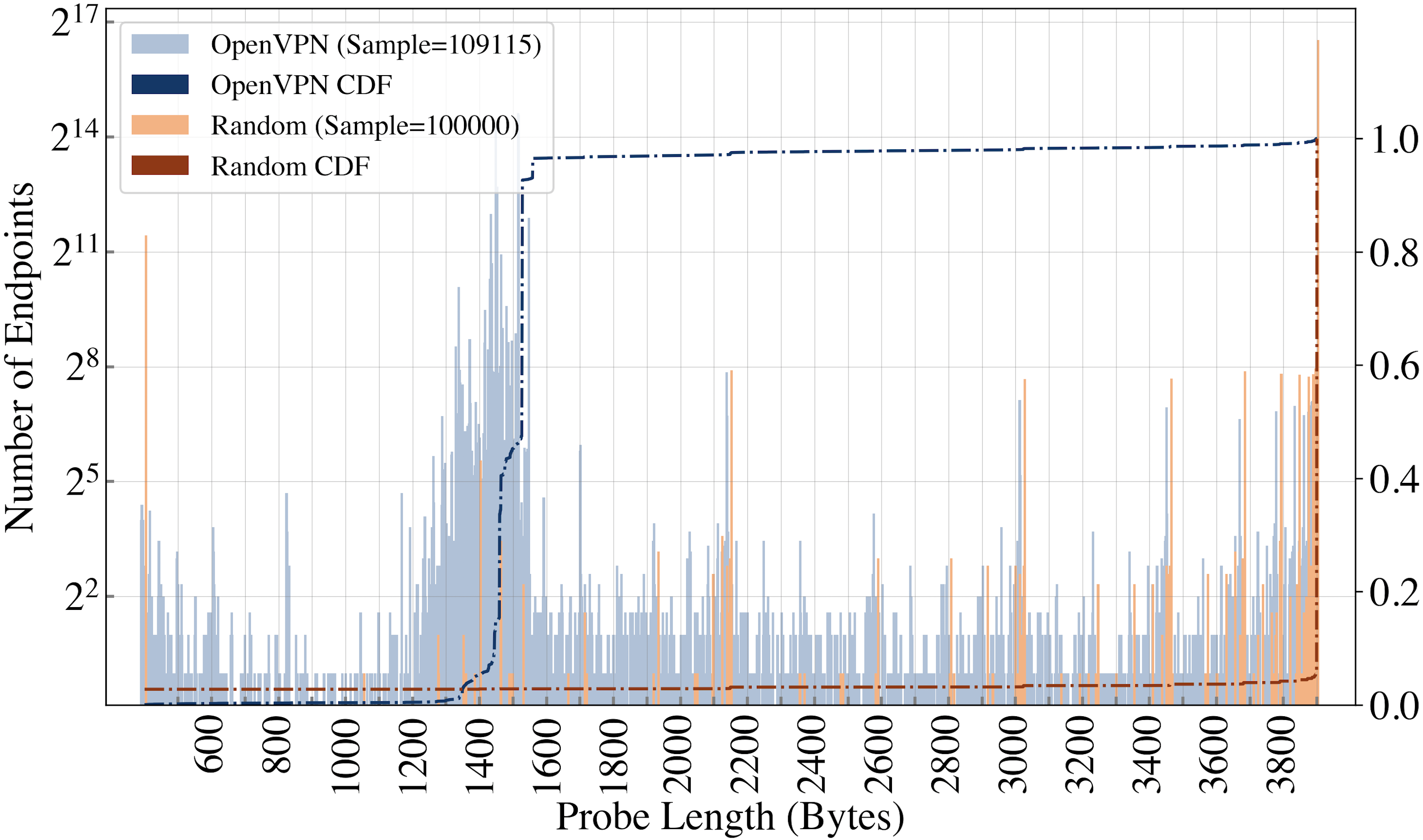}
\caption{\textbf{RST thresholds for OpenVPN and random endpoints.}}
\label{fig:rst-threshold}
\vspace{-15pt}
\end{figure}

We also design probes leveraging the way a Linux server closes a TCP connection. When a TCP connection terminates, the operating systems at both ends typically complete a FIN 4-way handshake. However, previous work has found that if a connection is closed with unread bytes in buffer, Linux will send a RST packet~\cite{Frolov-probe-resistant}. A server's ``RST Threshold'' is defined as the minimum number of bytes needed to send to the server to trigger a RST. We determine the RST threshold distribution for both \textit{ZMap Set} and \textit{Censys Set}. As shown in Figure~\ref{fig:rst-threshold}, the vast majority of OpenVPN servers have a RST threshold around 1550-1660 bytes, corresponding to buffers allocated with typical MTU configurations. In contrast, over 97\% of random ZMap endpoints have a RST threshold less than 500 or greater than 4000. We therefore construct an additional probe with 2,000 random bytes, which we expect over 98\% of legitimate OpenVPN servers and less than 3\% of random servers to respond to with RST packets.

\paragraph{Complication from Port Sharing} OpenVPN provides native support for another application to share the same port. This is accomplished by checking whether the first incoming packet has a valid OpenVPN-conforming length field. If not, the OpenVPN server will forward the packet to the other service sharing the port. This means that most of our additional probes will be forwarded to and responded by the other application due to invalid packet length. To account for this, we observe that when an OpenVPN shares a port, it is usually shared with a HTTP, TLS, or SSH service. Thus, we send probes targeting these three protocols after our base probes, and we stop further probing if we get an explicit response for any of these probes. 

It is worth noting that the majority of ``typical'' HTTP, TLS, and SSH servers have already been filtered out by our base probes, so endpoints that respond at this stage are likely sharing the port with another service, thus warranting manual analysis (\eg, checking TLS certificate). While these three services are what we commonly observed, there may be instances where other services are running along with OpenVPN. This could lead to false negatives.

Table~\ref{tab:probes} lists all probes and the expected behaviors from an OpenVPN server. An evaluation process is shown in Appendix Figure~\ref{fig:probe-tree}.

\subsection{Constructing \textit{Filters} and \textit{Probers}}

Our \textit{Filter} performs both opcode and ACK-based fingerprinting, flagging a flow if at least one fingerprint matches. This is because the opcode and ACK fingerprints are designed to be complementary: both are effective against vanilla OpenVPN and they each target a specific subset of obfuscations. The former works against XOR-based obfuscations that work like Vigenère ciphers, \ie they always encrypt the same plaintext opcodes at the same position to the same ciphertext bytes. The latter targets tunneling-based obfuscation that lacks random padding and preserves the 1:1 correspondence between the original and obfuscated packet streams. Combining the two features maximizes our fingerprinting coverage, as we discovered that even within the same provider, obfuscating strategies can vary a lot (\S~\ref{sec:result}). Table~\ref{tab:merit-results} in Appendix shows the effectiveness of each feature against each commercial VPN service we tested. Following \textit{Filter}'s result, the \textit{Prober} performs the active probing scheme to further lower potential false positives.

We implement the \textit{Filter} in Zeek~\cite{Zeek}, an open-source network monitoring tool. We note that the evaluation processes for opcode and ACK-based fingerprinting are quite simple: both only require several dozen integer comparisons (limited by the observation window) while maintaining a small number of per-flow states. We implement the \textit{Prober} in Nim~\cite{nim-lang}. We believe that both components can be easily deployed by any ISP or censor.


\section{Fine-tuning for Deployment}
\label{sec:finetune}
So far, we have described features that render OpenVPN vulnerable to fingerprinting. We still need to quantify detection thresholds (\eg ACK fingerprints) for implementation. Furthermore, there are metrics that can affect the system performance, such as packet loss or observation window choice. We seek to fine-tune our system by quantifying these parameters.

We use two datasets here. \textbf{ISP Dataset}: we collected a snapshot of network traffic going through a server installed within \textit{Merit}. Over 45 minutes on July 28, 2021, we sampled 1/30 of all flows passing through the server, resulting in 461~GB of traffic that corresponds to 221,534 flows with full packet payloads. Refer to \S~\ref{sec:ethics} for details on how this data was handled to limit privacy risks. \textbf{VPN Dataset}: we collected traces from 20 commercial VPN providers as well as 2 self-hosted OpenVPN services (Streisand, OpenVPN Access Server) following the automated process described in Section~\ref{sec:setup}. Note the 20 VPN providers do not overlap with the providers used in evaluation. For each provider, we repeated the trace collection process 50 times each in TCP and UDP mode, resulting in a 7.65~GB dataset comprised of 2,200 vanilla OpenVPN traces.

\subsection{ACK Fingerprint Thresholds}
\label{ACK-Threshold}

\begin{figure}[t]
\centering
 \includegraphics[width=\columnwidth,keepaspectratio]{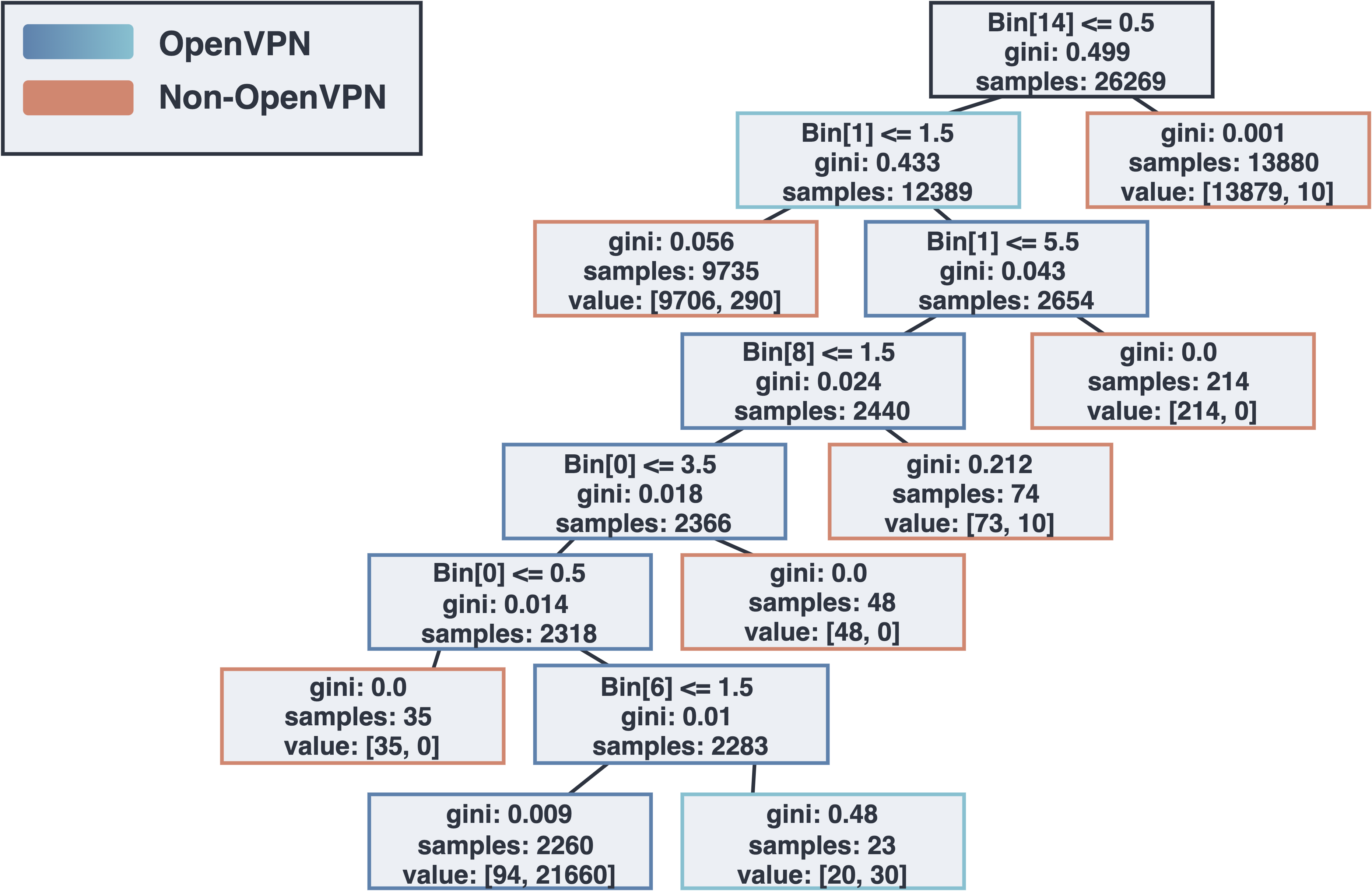}
\caption{\textbf{ACK fingerprint DT from the ISP and VPN datasets.}} 
\label{fig:ack-dt}
\vspace{-5pt}
\end{figure}
\begin{table}[]
\footnotesize
\begin{tabular}{p{0.45\columnwidth} p{0.45\columnwidth}}
\toprule
Bin Number     & Threshold \\ 

\midrule
1 & $1 \leq Bin[1] \leq 3$ \\ 
2 & $2 \leq Bin[2] \leq 5$\\ 
For $i$ in range $[3, 5]$ & $Bin[i] \leq 5$\\ 
For $i$ in range $[6, N/10]$ & $Bin[i] \leq 1$\\
\bottomrule
\end{tabular}
\caption{\textbf{Set of thresholds for ACK Filtering.}} 
\label{tab:ack-fin}
\vspace{-15pt}
\end{table}

We quantify the exact ACK fingerprint based on the \textit{ISP} and \textit{VPN Dataset}. We only include flows with at least 150 data packets (15 bins), which leaves us with 24,069 ISP flows and 2,200 VPN flows. A classification decision tree is constructed based on the two labeled sets with weights applied to account for the imbalanced data size. Figure~\ref{fig:ack-dt} shows the constructed tree (depth and leaf limited, a complete graph can be found in Appendix Figure~\ref{fig:ack-dt-complete}). The ACK fingerprint is a sequence of thresholds based on the derived decision tree, as shown in Table~\ref{tab:ack-fin}. ($Bin[i]$ refers to the number of ACK-size packets for $i^{th}$ Bin.)

\subsection{Choice of Observation Window N}
\label{sec:Observation-Window}

Previous works attempt to identify VPN traffic only \textit{after} the flow terminates, making use of aggregated statistics such as connection duration~\cite{Deep-Packets, VPN-nonVPN, perceptron-nn-VPN}. However, detecting disallowed traffic only after the flow is finished may be of limited interest to a real-world censor~\cite{On-the-fly}. We therefore have two objectives for our \textit{Filter}: to reduce probing targets by being as selective as possible, \textit{and} to detect OpenVPN as soon as possible within a flow.

Inspired by~\cite{On-the-fly, detect-obfuscation}, we consider the windowing strategy of limiting the inspection to only the first N data packets of a flow. We tested N from [10, 20, 30, ..... 200] on the \textit{ISP} and \textit{VPN Dataset}. As shown in Figure~\ref{fig:window-loss} (a), the number of ISP flows that are flagged by the \textit{Filter} declines from over 62,000 to 322 as we increase the observation window. However, we note that a window size of 100 packets has already achieved a precision within 2\% of the best performing (200 packets). 

\paragraph{Detection Speed and Potential Impact on Blocking} A smaller window size can sever a connection at an earlier stage, thereby reducing transfer of data to a censored endpoint, while a more conservative windowing strategy excels at accuracy. In our deployment, we use 100 packets as the window size to balance detection speed and accuracy. To put this choice into perspective, we note that the Great Firewall of China (GFW) was previously observed to send confirmation probes to suspected endpoints in 15-minute intervals, and it has only recently moved to near real-time operation~\cite{Ensafi-active-probing}. Recent work on how it detects Shadowsocks shows the median delay between the beginning of a connection and probing is about a minute, with probes being replayed for up to 47 times for confirmation~\cite{how-china-detect-shadowsocks}. In comparison, our deployment with a window size of 100 packets gives a median time of 7.9 seconds for the filter to flag an OpenVPN connection. We believe even with this delay, our system is still useful for censors who are interested in blocking OpenVPN connections. In addition, we note that a motivated adversary can further optimize this delay and speed the detection by tuning window size and probing rate, but with some potential loss of accuracy.

\subsection{Effects of Packet Loss}
\label{sec:loss}
We investigate the effects of packet loss on the performance of the \textit{Filter}. An adversary analyzing traffic on a busy network needs to keep pace with the packet arrival rate, or otherwise packet drops will start to occur due to a CPU bottleneck. For the opcode and ACK fingerprinting, we need to inspect the raw contents of each reassembled packet until the observation window is reached, for all flows.
This means all traffic must be passed on to Zeek's scripting layer, which may lead to a CPU bottleneck. In addition, the Network Interface Card (NIC) may also become an upstream bottleneck and lead to end-to-end packet loss. We therefore explore what to expect from the \textit{Filter} when packet loss is inevitable. 

\begin{figure}[t]
\centering
 \includegraphics[width=\columnwidth,keepaspectratio]{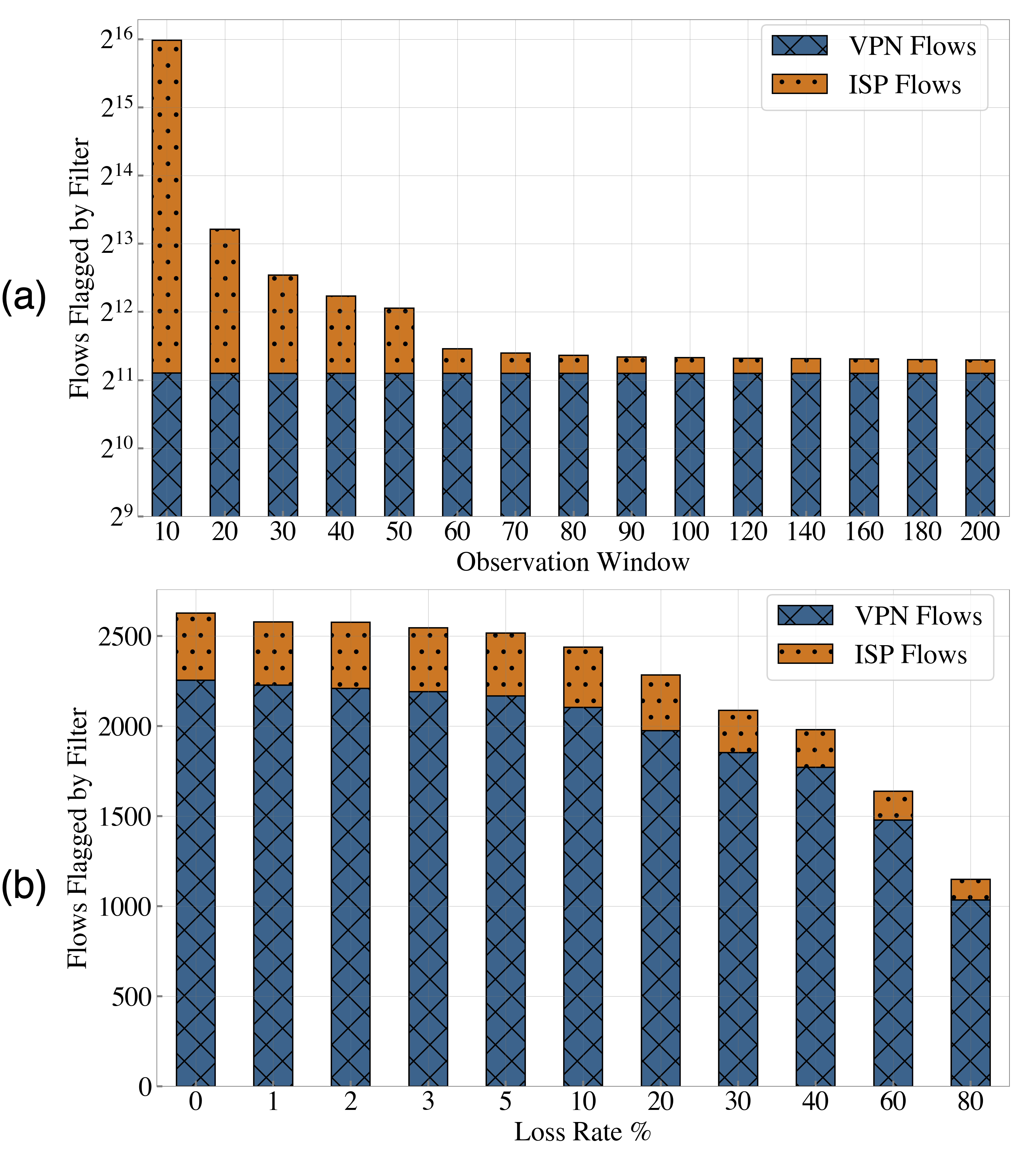}
\caption{\textbf{Effects of (a) observation window and (b) average packet loss rate on \textit{Filter}'s performance}}
\label{fig:window-loss}
\vspace{-15pt}
\end{figure}

We configure Zeek to ignore events signaling new packets with different probabilities in order to simulate random packet loss. We test loss rates from 1\% up to 80\%. The experiment is repeated three times, and the average result is reported in Figure~\ref{fig:window-loss} (b). We find that packet loss starts to have noticeable effects on the \textit{Filter}'s outputs once the loss rate surpasses 10\%. Notably, when packet loss starts to affect the detection accuracy, both opcode and ACK fingerprint vulnerabilities always produce \textit{underblocking} instead of \textit{overblocking}, which is favoured by real-world censors~\cite{sok-obfs}. Still, in order to minimize the effects of packet loss, we always configure the Monitoring Station to sample flows with a given rate (adjusted with CPU resources and traffic volume) for our online evaluation.
\looseness=-1

\subsection{Server Churn for Asynchronous Probing}
\label{sec:churn}
After the \textit{Filter} generates a list of probing targets, the \textit{Prober} can either send probes synchronously as soon as a target is emitted, or asynchronously, waiting for a pre-configured interval before sending probes to targets in batches. Sending probes synchronously has the advantage of obtaining the most accurate results before the server IP is churned. However, this requires the probing system to be online the whole time. In contrast, sending probes in batches is more efficient and easier to manage, but the server IP may be churned if the interval between the filtering phase and the probing phase is excessively long. We explore a probing frequency that achieves efficiency and accounts for possible server IP churn. To do this, we monitor the 180,858 known OpenVPN servers from the \textit{Censys Set} described in \ref{active-probing}. Starting from August 2nd, 18:00 EDT, we probe the servers every 3 hours for a week and record their responses.

\begin{figure}[t]
\centering
 \includegraphics[width=\columnwidth,keepaspectratio]{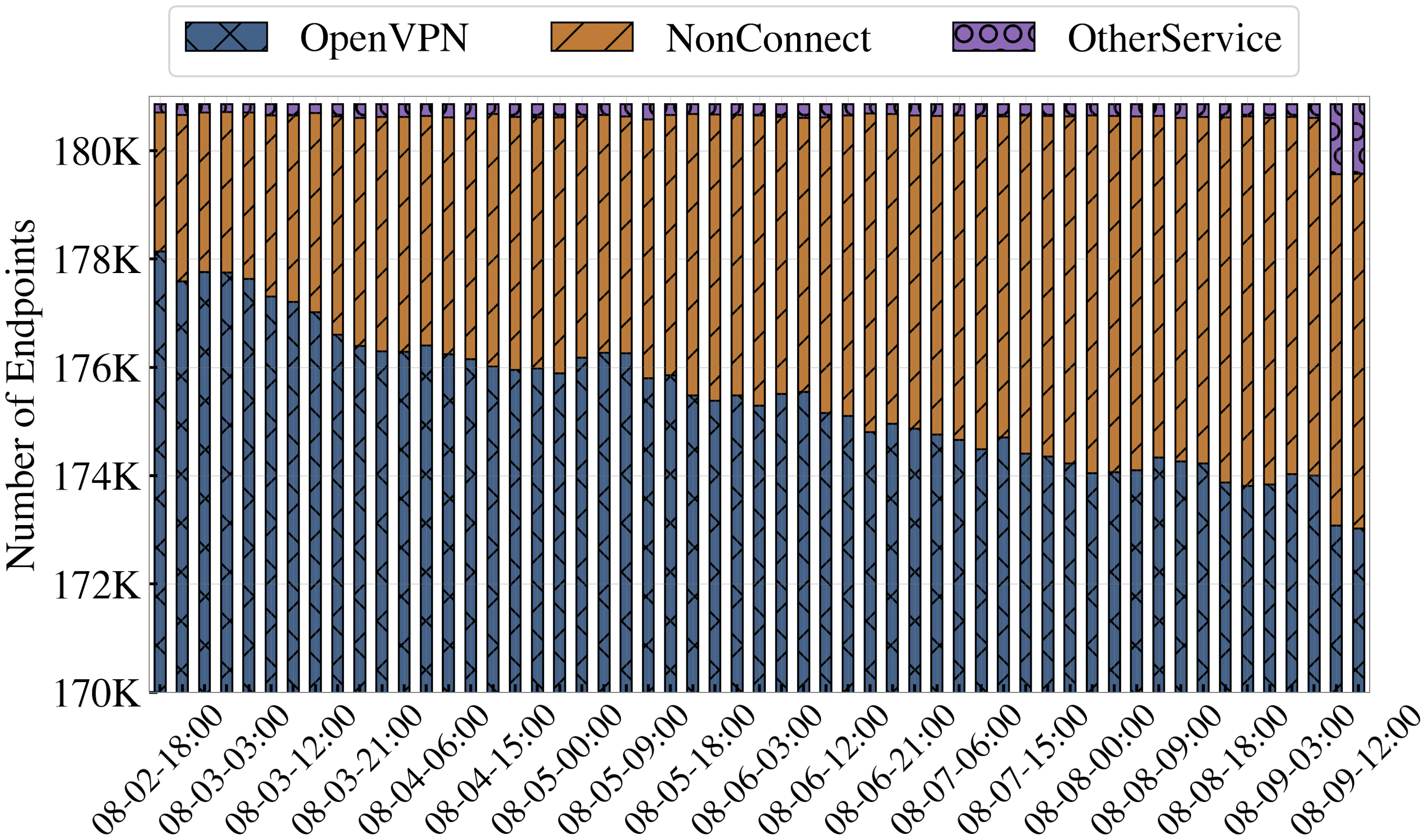}
\caption{\textbf{OpenVPN server churns over time.}}
\label{fig:churn}
\vspace{-15pt}
\end{figure}

As shown in Figure~\ref{fig:churn}, even after a week, only 2.39\% of OpenVPN servers either are not in the listening state or have been replaced by a different service. This suggests that the majority of OpenVPN servers are not churned frequently. In our online evaluation, we choose to probe targets in batches on a daily basis to balance between efficiency and potential IP churn. Based on the result of this test, approximately 0.9\% of servers may be churned within 24 hours.

\subsection{Probe UDP and Obfuscated OpenVPN Servers}
\label{sec:probe-subnet}
The active probing scheme in the previous section primarily targets vanilla OpenVPN TCP servers, as it exploits the header length field that is unique to TCP mode that requires packetization. In addition, it works effectively against XOR-obfuscated servers because the length field is prefixed \textit{after} the XOR encryption is applied to an OpenVPN packet. This construction allows us to probe XOR-obfuscated servers in the same way as if they had no obfuscation at all.

For UDP or other obfuscated servers, our probes are no longer effective because the length field is either not present (UDP) or encrypted (tunnel-based obfuscation). However, a critical observation is that most commercial VPN providers usually offer vanilla TCP servers along with UDP and/or obfuscated variants. This is expected as commercial VPN providers attempt to optimize their VPN's performance as well as reliability, since tunnel-based obfuscation adds overhead and UDP traffic may encounter more problems than TCP in a firewalled network. Furthermore, the vanilla TCP service is often co-located with the UDP or obfuscated OpenVPN services, presumably due to lower hosting and maintenance cost. They could be on the same host by listening on different ports, or they could be located in adjacent IPs in the same VPN provider subnet. In other words, probing adjacent netblocks of a suspected UDP or obfuscated endpoint may reveal nearby vanilla TCP servers, whose existence corroborates the \textit{Filter} results. For our \textit{Prober} deployment on two dedicated measurement machines, we limit our probing to the /29 subnet the target IP belongs to over all TCP ports. This specific subnet size is chosen primarily due to probing resources limitation, and a more well-resourced adversary may expand the probing to larger subnets. With only two measurement machines, the parallelized /29 \textit{Prober} is able to probe targets generated by a \textit{Filter} monitoring a 5~Gbps network interface.

\section{Real-world Deployment Setup}
\label{sec:setup}

We set out to explore if an ISP or censor can fingerprint OpenVPN connections at scale, without significant collateral damage. Adopting the viewpoint of an adversarial ISP, we deploy our framework inside \textit{Merit}, as shown in Figure~\ref{fig:deployment}. Our evaluation is two-fold: we generate \textit{control} vanilla and obfuscated flows with commercial VPN providers and attempt to identify them as a network intermediary; we also process other traffic passing through our Monitoring Station in order to estimate the false positive rate of our framework.


We set up our framework on a 16-core server (Monitoring Station) inside \textit{Merit} with two mirroring interfaces that have an aggregated 20~Gbps bandwidth. Due to the large traffic volume, we optimize our deployment with PF\_RING~\cite{pf-ring-zc} in order to improve the packet processing speed. We employ PF\_RING in zero-copy mode and spread the traffic load across a Zeek cluster of 15 workers. Nonetheless, due to limited CPU resources, we only sample 12.5\% of all TCP and UDP flows arriving at the network interfaces in order to minimize the effect of packet loss. The sampling is based on IP pairs so that all bi-directional traffic of a flow will be selected/dropped together. With these settings, we are able to operate with an end-to-end packet loss rate under 3\%. Even though we process only a fraction of all traffic, our \textit{Filter} still handles over 15 Terabytes of traffic from over 2 billion flows on an average day on a single server. In addition, processing all traffic without sampling is feasible through parallelism or using faster CPUs.
\looseness=-1

Next, we set up \textit{Probers} on two dedicated measurement machines, each provisioned with 10 IPv4 and 1 IPv6 addresses. By the end of each day during the evaluation, the \textit{Probers} fetch filtering logs from the Monitoring Station. For each target, we run a Masscan~\cite{masscan} to the /29 subnet the IP belongs to over all TCP ports (1-65535). We follow up each discovered open port by running our probing scheme, and endpoints confirmed through probing are recorded for manual analysis.

To select VPN services for evaluation, we first generate a list of ``top'' VPN services ranked by popularity. We combine 80 providers, most of which are paid premium VPN services, from top VPN recommendation sites based on previous work~\cite{vpnalyzer-ndss}, listed in Appendix Table~\ref{tab:recommendation}. Next, we visit the websites of these VPN providers searching for ``Obfuscation'', ``Stealth'', or ``Camouflage Mode'' \etc, and include providers that offer at least one obfuscated VPN configuration. 
In total, we find 24 providers offering obfuscated services. We test all obfuscation configurations if more than one is offered as well as vanilla OpenVPN for each provider. If TCP and UDP modes are both available, we test them separately. In total, we have 81 configurations, 41 of which are obfuscated ones.

We configure the Client Station inside \textit{Merit} to act as a VPN client. Both upstream and downstream traffic of the Client Station go through the router that mirrors traffic to the Monitoring Station. In addition, we exclude this server from our random sampling so that \textit{all} traffic to/from this server will be analyzed. On the client, we run an automated script to generate \textit{control} traffic for our evaluation. For each iteration, we start the VPN client application and connect to the ``default / recommended'' server using Pywinauto~\cite{Pywinauto}. After a random wait of 20 to 180 seconds, we confirm that the VPN tunnel is active and generate random browsing traffic with Selenium~\cite{selenium} by sending requests to a random website from the Alexa top 500. Finally, we disconnect from the VPN server and wait for 180 seconds before proceeding to the next iteration. For each VPN configuration, we repeat the process 50 times and collect packet captures for reference.
\section{Evaluation \& Findings}
\label{sec:result}

\begin{table}[]
\footnotesize
\begin{tabular}{p{0.19\columnwidth}p{0.4\columnwidth}p{0.3\columnwidth}}
\toprule
\midrule
            Control Flows       & Overall Recall              & 3141/4120 (76.24\%) \\  
                     & Filter Recall          & 3635/4020 (90.42\%) \\  
                     & Prober Recall          & 3186/3635 (87.65\%) \\ 
                     & Vanilla Recall        & 1718/2000 (85.90\%) \\ 
                     &Obfuscated Recall      & 1468/2020 (72.67\%) \\
\midrule
            All Flows         & Flow Count & 23183039736  \\  
            & Bytes Processed & 124.67 Terabyte \\ 
            & Flows $\geq$ Observation Window         & 10070994  \\ 
            & Filter Outputs  & 75850 \\  
            & Probing Outputs & 3638  \\ 
            \midrule
            & Confirmed OpenVPN Flows & 3245 \\
            & Remaining Unclassified & 393 (0.0039\%)\%  \\ 
        
\midrule
\bottomrule
\end{tabular}
\caption{\textbf{High-level evaluation statistics on \Merit.}}
\label{tab:control-result}
\vspace{-15pt}
\end{table}

We started the evaluation on August 13, 2021, and kept the monitor running for a week until August 20. Table~\ref{tab:control-result} contains high-level statistics of the evaluation. A more detailed result that breaks down statistics by each \textit{control} VPN configuration can be found in Appendix Table~\ref{tab:merit-results}.

\subsection{Results for \textit{control} VPN flows}

Overall, we are able to identify 1,718 out of 2,000 vanilla flows, corresponding to 39 out of 40 unique configurations. This suggests the majority of OpenVPN traffic and servers are vulnerable to passive filtering and active probing, respectively. The few exceptions correspond to VPN providers that only offer UDP-based services or hide their servers behind IDS~\cite{cryptostorm-port}, which thwarts our probing attempts. Surprisingly, we also identify over two-thirds of all obfuscated flows, corresponding to 34 out of 41 obfuscated configurations. This result is mostly due to obfuscated services using OpenVPN as their backbone protocol and insufficient obfuscation failing to mask OpenVPN's fingerprints. Alarmingly, out of the ``top 10'' VPN providers ranked by \textit{top10vpn.com}~\cite{top-10-vpn}, eight provide obfuscation services of some sort, suggesting that being undetectable is within the providers' threat model for their clients. Yet, \textit{all} of them are flagged as suspect flows due to either insufficient encryption (Opcode) or insufficient obfuscation over packet length (ACK). Considering that these obfuscated VPN services usually claim to be ``undetectable'' or claim that the obfuscation ``keeps you out of trouble''~\cite{Boleh-Obfuscated, TorGuard-Obfuscated}, this result is alarming as users who use these services may have a false sense of privacy and ``unobservability''.

\begin{figure*}[t]
\centering
 \includegraphics[width=2\columnwidth,keepaspectratio]{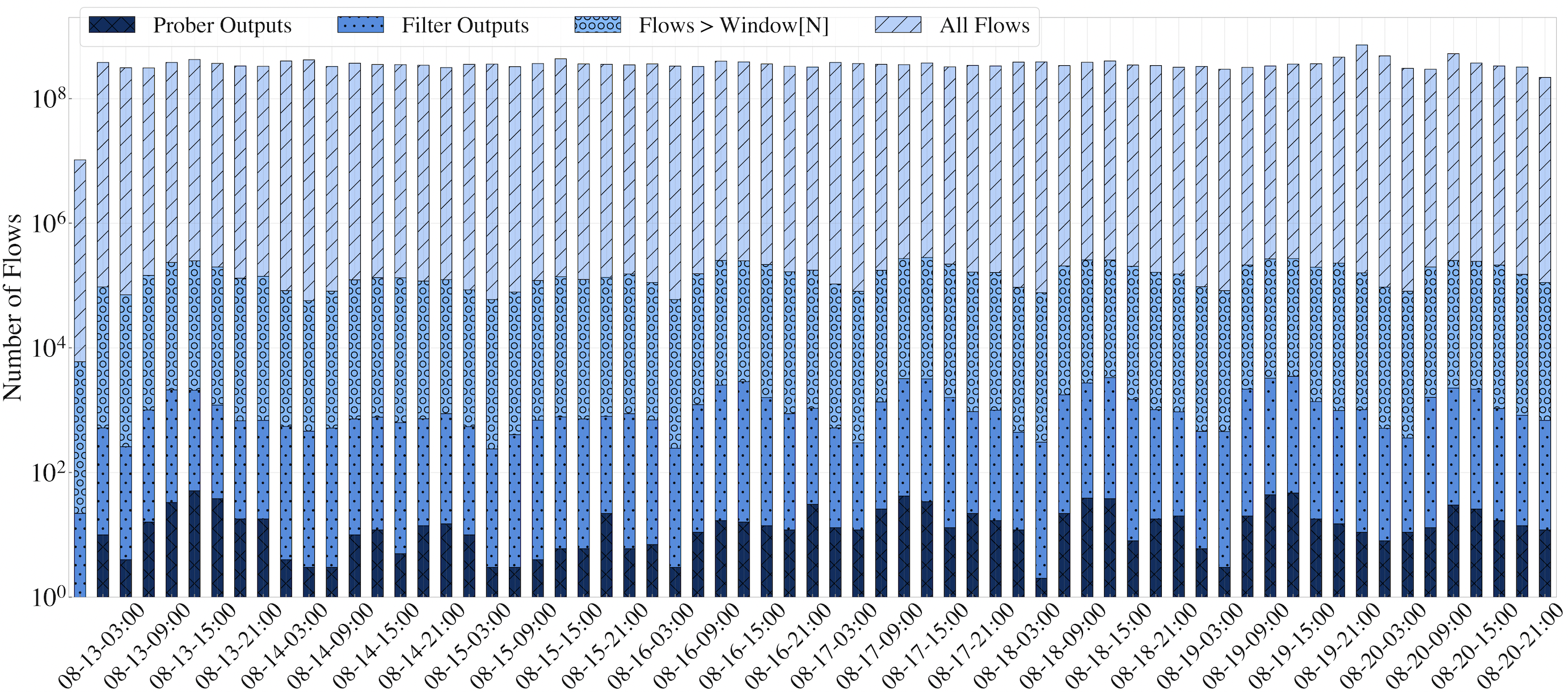}
\caption{\textbf{\Merit evaluation results over days, excluding Control VPN flows.}}
\label{fig:eval}
\vspace{-15pt}
\end{figure*}

\myparagraph{4 out of the ``top 5'' VPN providers use XOR-based obfuscation, which is easily fingerprintable.} We find that among the ``top 5'' VPN providers~\cite{top-10-vpn}, four offer obfuscated services, all of which nonetheless are flagged as OpenVPN flows by our \textit{Filter} over 90\% of the time. A closer look at the raw packet capture suggests that \textit{all} of them employ obfuscations that are almost identical to the unofficial XOR patch, thereby making them vulnerable to fingerprinting. Alarmingly, the XOR-based obfuscation---despite being rejected by OpenVPN developers~\cite{Tunnelblick-XOR}---appears to be a major obfuscation strategy adopted by the majority of VPN providers we test, who often praise the patch for its simplicity and low overhead. Although the patch can bypass some of the most basic filters adopted by existing open-source DPI tools, we have demonstrated that even a slightly more sophisticated filter will be able to reliably and accurately detect them. 

\myparagraph{Wrapping OpenVPN inside encrypted tunnels is a popular obfuscation strategy, yet some flows are still recognizable due to a lack of random padding.} Another popular class of obfuscation strategies is tunnel-based, which wraps OpenVPN traffic inside an encrypted tunnel to frustrate any analysis over packet payloads. Examples include Stunnel, SSH tunnel, Shadowsocks, obfs\{2/3/4\}, and V2Ray(VMess). Overall, we find 20 obfuscated configurations deployed by 14 VPN providers that are tunnel-based. However, most of these tunnels do not add random padding to the payload being tunneled, with the only exceptions being obfs4 and VMess which can draw packet sizes from certain distributions. Among the 20 tunnel-based obfuscated services, only three of them deploy obfs4 and only one deploys VMess, leaving the remaining 16 vulnerable to ACK fingerprinting. We note that this does not mean these tunneling tools do not work, but rather that protection against traffic analysis is not among the design goals. For example, the threat model of obfs3, which is deployed by Perfect Privacy VPN, states that the obfuscator ``does not try to protect against non-content protocol fingerprints, like the packet size or timing''~\cite{Obfs3-threatmodel}. Yet, we have demonstrated that for applications with distinct signature over packet length, such as OpenVPN, even the simplest threshold-based detection can identify them with reasonable accuracy.
\looseness=-1

\myparagraph{UDP and obfuscated servers often share infrastructure with vanilla TCP servers, leaving them ``guilty by association''.}
We discover that the majority of UDP and obfuscated OpenVPN services are co-located with vanilla TCP servers. For example, TorGuard hosts vanilla and stunnel-obfuscated OpenVPN instances on the same host but different ports, whereas Perfect Privacy hosts them in neighboring IPs (*.*.193.26 for vanilla, *.*.193.27 for Stunnel, *.*.193.28 for SSH, and *.*.193.29 for obfs3). We find that for 34 out of 41 obfuscated services, at least one vanilla OpenVPN TCP server can be found within the server's /29 subnet. Similarly, we were able to actively probe 18 out of 20 UDP configurations due to their co-location with TCP servers.
In addition, we also find five providers sharing infrastructures used by their obfuscated services. For example, one IP (23.95.*.*) hosts Cryptostorm's SSH-obfuscated service as well as SecureVPN's vanilla servers.
This result is only a lower bound as we did not connect to every single server available from each provider. Obfuscated services using shared infrastructures may be easier for adversaries to identify and block.

On the positive side, some deployed services successfully evade our detection. Some providers deploy randomizers such as obfs4, v2ray, or proprietary protocols with random padding, which stopped us at the filtering stage (\eg Tunnelbear). In addition, some providers deploy their obfuscated servers behind a firewall or IDS, which would respond with SYN-ACKs to every arriving SYN packet on almost all TCP ports~\cite{LZR, Frolov-probe-resistant}. Since we limit probe targets to 2,000 open ports per IP for practical considerations, this leads to false negatives when none of the probes hit an OpenVPN-listening port. Moreover, some VPN providers do not host vanilla OpenVPN TCP servers at all, such as VyprVPN, which currently only supports UDP as transport. For these providers, even though our \textit{Filter} flags their flows as suspected OpenVPN, we were not able to confirm with subsequent probing. Finally, some providers host UDP or obfuscated servers outside the /29 probing range of the vanilla TCP ones, and we miss them due to probing resource constraints.


\subsection{Results for all flows}

Figure~\ref{fig:eval} shows an hour-level breakdown of the evaluation statistics, excluding \textit{control} flows.
Overall, both the \textit{Filter} and \textit{Prober} are able to reduce the number of suspected flows by several orders of magnitude, which when combined flagged 3,638 flows as OpenVPN connections. We manually analyze these flows to confirm our detection results.


Among the 3,638 flows, the destination servers for 469 of them respond to our \textit{Base Probe \#1} with an explicit server reset, indicating the presence of a legitimate OpenVPN server not configured with HMAC protection. For the remaining 3,169 flows, we first noticed that 2,580 of them are between a single IP pair. Based on our log, the client initiates a connection every 4 minutes to the server on port 1194 (assigned to OpenVPN). Reverse DNS lookup associates the client IP with the ``lib-locker'' subdomain under a private university in the US. Furthermore, the server runs a TLS service listening on port 443, which sends a certificate belonging to a smart locker company with subject and issuer CN as ``vpn.\_COMPANY\_.com''. Based on these evidence, we believe the captured flows correspond to the secure communications between a deployed smart locker and the infrastructure that controls it. This also suggests that the fingerprintability of OpenVPN may not only be a problem concerning censorship circumvention, but it may also be used for reconnaissance to identify and target IoT devices that communicate to their servers over an OpenVPN channel. Finally, we attempt to further characterize the remaining 589 flows based on circumstantial evidence about the destination endpoint.

\paragraph{Co-location with TLS} In practice, TLS is the most common application we have seen that is co-located with an OpenVPN instance. For each of the remaining flows, we probe its destination endpoint with a TLS Client Hello and analyze the certificate and web page returned. Endpoints of 40 flows return certificates whose subject or issuer CN suggest VPN activity, such as \texttt{*.vpn.ipvanish.com}, \texttt{*.vpn.wlvpn.com}, \texttt{*.virtualshield.org}, and \texttt{OpenVPN Web CA}. In addition, 16 endpoints serve OpenVPN web interfaces over TLS.

\paragraph{WHOIS, DNS PTR, ISP Name} We look up the WHOIS and DNS PTR records of the destination endpoints. 11 server IPs of 41 flows contain WHOIS records that can be linked back to a VPN provider, such as \texttt{protonvpn-*}, \texttt{PRIVADO-*}, and \texttt{secureconnectivity-*}. In addition, 2 servers have DNS PTR records as \texttt{*.strong.blackoakcomputers.com} and \texttt{fosvpncluster.fos.*.com}.

\paragraph{IP Context Service} Several online platforms claim to offer VPN IP database or IP context services. We found 124 flows that can be linked to a commercial VPN server IP by the lookup service hosted on \texttt{spur.us}. However, these services do not disclose their specific methodology and their accuracy has not been systematically evaluated.

Our 7-day evaluation flagged 3,638 flows that are identified as ``OpenVPN'' from over 10 million flows that exceed our observation window. Among these, we are able find evidence that supports our detection result for 3,245 flows. The majority of the remaining 393 flows have server IPs belonging to cloud hosting services, and we are not able to further classify them. Conservatively, we can upper bound the false positive rate to 0.0039\%, which is three orders of magnitude lower than previous ML-based approaches (1.4\%-5.5\%)~\cite{VPN-Comparison, perceptron-nn-VPN, LSTM-Attention}

\section{Discussion and Mitigations}
\label{sec:discussion}



ISPs and government censors are motivated to detect OpenVPN flows in order to enforce traffic policies and information controls. We demonstrate that tracking and blocking the use of OpenVPN, even with most deployed obfuscation methods, is practical at scale and with minimal collateral damage. We note that many VPN providers’ claims that their obfuscated services are unobservable appear to be misleading and potentially dangerous, especially to users from countries where personal VPN usage is illegal. In light of our findings,  users should \emph{not} expect complete unobservability even when connected to ``obfuscated’’ OpenVPN-based services.

Putting the human danger aside, the ease of fingerprinting makes OpenVPN more susceptible to throttling or blocking from ISPs and governments. Previous research suggests that some censors already use two-stage pipelines, which are highly similar to our deployment, to detect other protocols such as Tor or Shadowsocks~\cite{how-china-detect-tor, how-china-detect-shadowsocks}. These adversaries can quickly adapt such infrastructure to detect OpenVPN traffic by simply adding protocol-specific fingerprints and probes. Furthermore, while we focus on OpenVPN due to its overwhelming popularity among commercial VPNs, it is possible to extend our two-stage framework to other VPN protocols (\eg, WireGuard and StrongSwan) by analyzing their communication patterns and server behaviors. Governments can also quickly adopt these fingerprints to track and block VPN usage during sensitive times, like political upheavals, when VPN connections are most vital to the free flow of information.

\paragraph{Short-Term Mitigation}

There are several defensive strategies to achieve near-term protection from the fingerprinting attack we describe. First, VPN providers offering both vanilla and obfuscated OpenVPN services should avoid co-locating them. Ideally, obfuscation servers should be well separated from OpenVPN instances in the network address space and operate as ``bridge servers'' that forward client traffic to VPN servers elsewhere. For example, Mullvad VPN offers a Shadowsocks-based obfuscator service as a dedicated bridge, separating the VPN servers from the obfuscation~\cite{Mullvad-shadowsocks}.

Second, VPN providers should switch from static to random padding for their obfuscated services. As we have shown, for protocols with a stable and distinctive handshake phase, even the most basic threshold-based detector is able to fingerprint them by packet sizes. Ideally, the obfuscation layer should be able to send zero-length packets (packets whose payloads are all padding) to break the one-to-one correspondence between the unobfuscated and obfuscated packet streams~\cite{wireguard-obfuscation-1}. Yet it is worth noting that previous work has shown that even fully randomized obfuscators (\eg, obfs4) can themselves be vulnerable to entropy-based fingerprinting attacks~\cite{detect-obfuscation}.

Third, we suggest that the OpenVPN developers follow recommendations from previous work with regard to how servers respond to failed handshake attempts. Servers closing failed connections immediately or in a predictable manner has enabled active probing attacks against a variety of other protocols~\cite{how-china-detect-shadowsocks, v2ray-weakness, Frolov-probe-resistant}. In response, these protocols have implemented either unlimited timeouts (reading from the buffer indefinitely) or diversified close behaviors (in which each server instance closes failed connections in a different manner).
\looseness=-1

\paragraph{Long-Term Defenses}

In the long term, we fear that the cat-and-mouse game between censors and circumvention tools, such as the Great Firewall and Tor, will occur in the VPN ecosystem as well, and developers and providers will have to adapt their obfuscation strategies to the evolving adversaries. We urge commercial VPN providers to adopt more standardized obfuscation solutions, such as Pluggable Transports~\cite{Pluggable-Transports-Project}, and to be more transparent about the techniques used by their obfuscated services. This transparency will help foster development of stronger obfuscation methods and encourage developers to design better techniques to overcome the progress of information control technologies.
Additional future work is needed to characterize the performance costs of different approaches to VPN obfuscation and to help users with varying threat models make appropriate trade-offs between performance and resilient unobservability.

\section{Conclusion}
\label{sec:conclusion}
We have demonstrated that OpenVPN, even with widely applied obfuscation techniques, can be reliably detected and blocked at-scale by network-based adversaries. Inspired by previous real-world censorship events, we designed a two-phase system that performs passive filtering followed by active probing to fingerprint OpenVPN flows. We evaluated the practicality of our approach in partnership with a mid-size ISP, and we were able to identify the majority of vanilla and obfuscated OpenVPN flows with only negligible false positives, which supports that the techniques we describe would be practical even for adversaries averse to collateral damage.

Users worldwide rely on VPNs to protect their security and privacy and to escape Internet censorship, yet the ease of fingerprinting OpenVPN traffic and the commodification of DPI technologies bring monitoring and blocking of popular VPN services within reach for almost any network operator. We propose several short-term mitigations that can help defend against these threats, but in the long term, we urge VPN providers to adopt more resilient and better standardized obfuscation approaches.

\section{Acknowledgement}
\label{sec:ack}

The authors are grateful to Matthew Wright for shepherding the paper, and to the anonymous reviewers for their constructive feedback. This material is based upon work supported by the National Science
Foundation under Grant No.1518888, 1823192, 2007741, 2042795, 2120400.

{\clubpenalty=0\widowpenalty=0
\bibliographystyle{abbrv}
\bibliography{paper}}
\appendix
\section{Appendix}
\label{sec:appendix}

\begin{figure}[h]
\centering
 \includegraphics[width=\columnwidth,keepaspectratio]{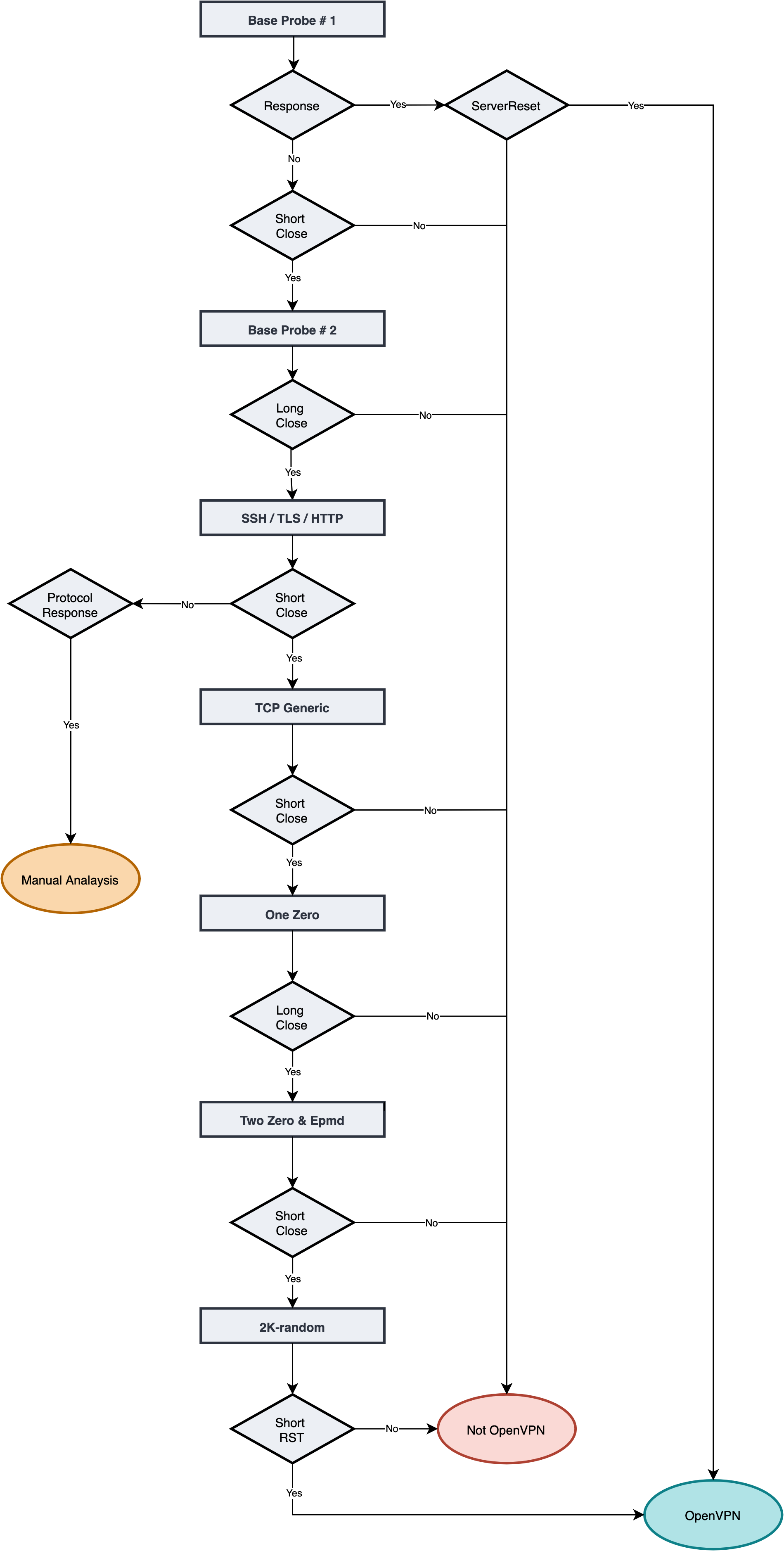}
\caption{\textbf{Evaluation Process for Active Server Fingerprinting.}}
\label{fig:probe-tree}
\vspace{-15pt}
\end{figure}
\begin{figure}[t]
\centering
 \includegraphics[width=\columnwidth,keepaspectratio]{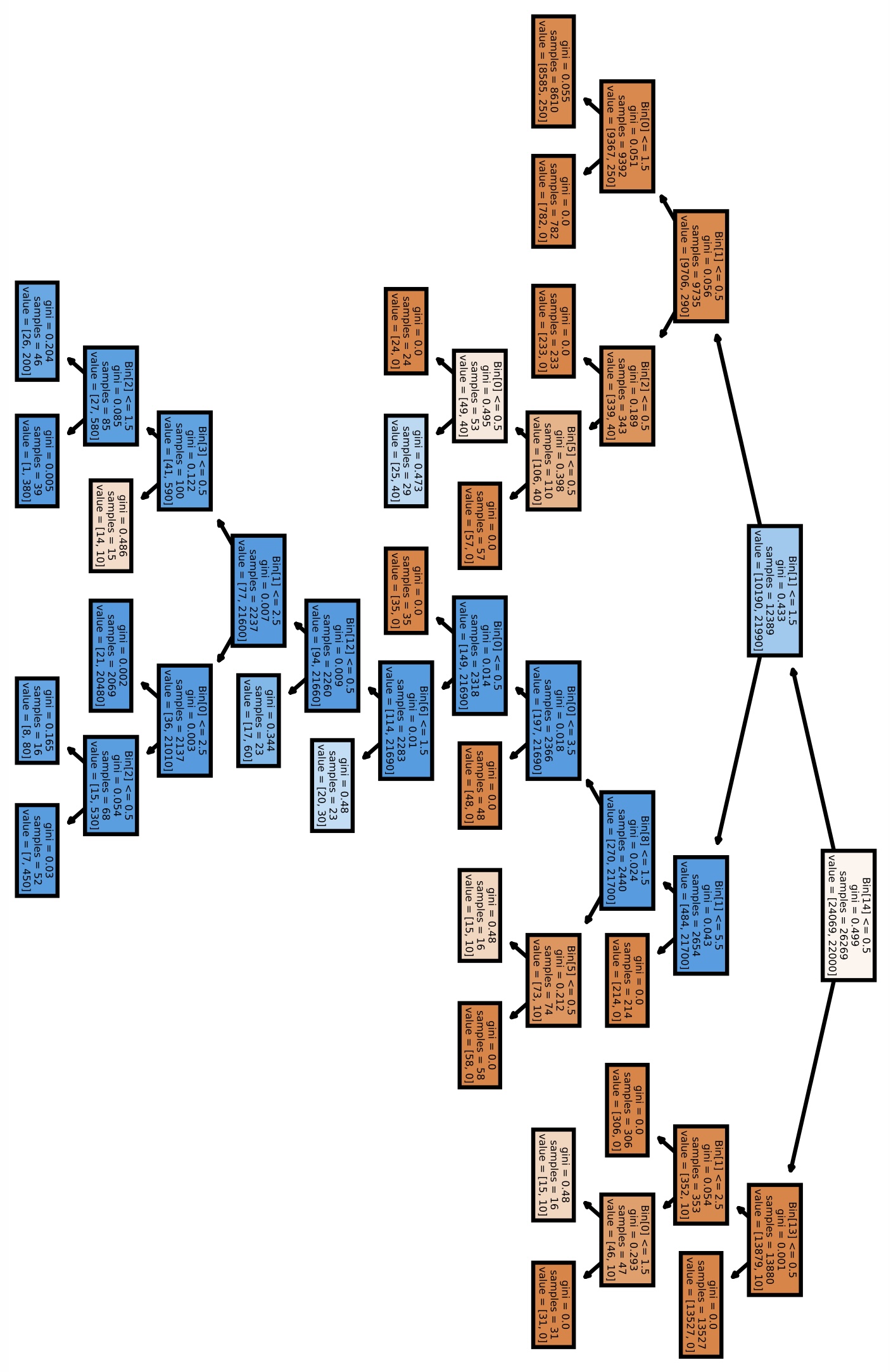}
\caption{\textbf{Decision tree derived from ISP and VPN datasets.}}
\label{fig:ack-dt-complete}
\end{figure}




\begin{table}[t]
    \scriptsize
    \centering
\begin{tabular*}{\columnwidth}{@{\extracolsep{\fill}}l}   
        \toprule
        {VPN Recommendation Sites Used}  \\
        \midrule
         https://www.security.org/vpn/best/	\\
        https://www.techradar.com/vpn/best-vpn	\\
        https://www.cnet.com/news/best-vpn/	\\
        https://www.tomsguide.com/best-picks/best-vpn	\\
        https://www.pcmag.com/picks/the-best-vpn-services	\\
        https://thebestvpn.com/	\\
        https://www.wired.co.uk/article/best-vpn	\\
        https://www.zdnet.com/article/best-vpn/	\\
        https://www.cloudwards.net/best-vpn/	\\
        https://www.internetsecurity.org/compare/usa	\\
        https://www.top10vpn.com/best-vpn-for-usa/v/d/?bsid=c33se1kw011	\\
        https://bestvaluevpn.com/usd/best-vpn/?utm_campaign=ggls-en-usa-gen	\\
        https://www.nytimes.com/wirecutter/reviews/best-vpn-service/	\\
        https://cybernews.com/best-vpn/	\\
        https://vpnoverview.com/best-vpn/top-5-best-vpn/	\\
        https://www.guru99.com/best-vpn-usa.html	\\
        https://www.crazyegg.com/blog/best-vpn-services/	\\
        https://www.forbes.com/advisor/business/software/best-vpn/	\\
        https://blog.flashrouters.com/vpn/	\\
        https://vpnpro.com/best-vpn-services/	\\
        https://bestvpn.org/best-vpns-for-the-usa/	\\
        https://www.safetydetectives.com/best-vpns (formerly thatoneprivacyguy)	\\
        https://www.tomsguide.com/best-picks/best-free-vpn	\\
        https://www.top50vpn.com/best-vpn	\\
        https://www.top10vpn.com/best-vpn/	\\
         \midrule
        \bottomrule
    \end{tabular*}
    \caption{\textbf{Recommendation Websites Used}}
    \label{tab:recommendation}
\end{table}

\definecolor{myblue}{rgb}{0.5,0.605,0.894}

\begin{table*}
    \tiny
    \centering
    \begin{tabular*}{2\columnwidth}{@{\extracolsep{\fill}}p{2.75cm}|p{1.3cm}p{2.2cm}|p{1.3cm}p{1cm}p{1cm}p{1.5cm}p{1cm}|p{1.2cm}}

        
         VPN Provider Name &	Transport & Variant	&  Collection Size & Filter & 	Filter Rate & Opcode/ACK & Prober  &	 Overall Rate \\
        
        \midrule
        
        \rowcolor{myblue}
AirVPN& TCP& SSL Tunnel& 50& 49& 0.98& 0/49& 49& 0.98 \\
\rowcolor{myblue}
AirVPN& TCP& SSH Tunnel& 50& 16& 0.32& 0/16& 16& 0.32 \\
AirVPN& TCP& Vanilla& 50& 48& 0.96& 48/48& 37& 0.74 \\
AirVPN& UDP& Vanilla& 50& 49& 0.98& 49/49& 49& 0.98 \\
\rowcolor{myblue}
Astrill& TCP& Proprietary& 50& 47& 0.94& 0/47& 9& 0.18 \\
\rowcolor{myblue}
Astrill& UDP& Proprietory/XOR*& 50& 50& 1& 50/6& 0& 0 \\
Astrill& UDP& Vanilla& 50& 49& 0.98& 49/7& 4& 0.08 \\
\rowcolor{myblue}
BolehVPN& TCP& XOR*& 50& 50& 1& 50/50& 50& 1 \\
\rowcolor{myblue}
BolehVPN& UDP& XOR*& 50& 49& 0.98& 49/46& 49& 0.98 \\
BolehVPN& TCP& Vanilla& 50& 50& 1& 50/50& 50& 1 \\
BolehVPN& UDP& Vanilla& 50& 50& 1& 50/49& 50& 1 \\
\rowcolor{myblue}
CactusVPN& TCP& Obfsproxy& 50& 0& 0& 0/0& 0& 0 \\
CactusVPN& TCP& Vanilla& 50& 50& 1& 48/50& 50& 1 \\
CactusVPN& UDP& Vanilla& 50& 50& 1& 50/50& 50& 1 \\
\rowcolor{myblue}
Cryptostorm& TCP& HTTPS Tunnel& 50& 48& 0.96& 0/48& 19& 0.38 \\
\rowcolor{myblue}
Cryptostorm& TCP& SSH Tunnel& 50& 27& 0.54& 0/27& 7& 0.14 \\
Cryptostorm& TCP& Vanilla& 50& 48& 0.96& 48/48& 13& 0.26 \\
Cryptostorm& UDP& Vanilla& 50& 49& 0.98& 49/49& 16& 0.32 \\
\rowcolor{myblue}
ExpressVPN& TCP/UDP& XOR*& 20& 20& 1& 20/4& 16& 0.8 \\
ExpressVPN& TCP& Vanilla& 50& 49& 0.98& 49/49& 29& 0.58 \\
ExpressVPN& UDP& Vanilla& 50& 50& 1& 50/50& 32& 0.64 \\
\rowcolor{myblue}
Hide.me& TCP& Vanilla& 50& 49& 0.98& 49/49& 49& 0.98 \\
\rowcolor{myblue}
Hide.me& UDP& Vanilla& 50& 45& 0.9& 45/45& 41& 0.82 \\
\rowcolor{myblue}
IPVanish& TCP& XOR*& 50& 49& 0.98& 49/49& 49& 0.98 \\
\rowcolor{myblue}
IPVanish& UDP& XOR*& 50& 50& 1& 50/50& 50& 1 \\
IPVanish& TCP& Vanilla& 50& 50& 1& 50/50& 50& 1 \\
IPVanish& UDP& Vanilla& 50& 47& 0.94& 47/47& 47& 0.94 \\
\rowcolor{myblue}
IVPN& TCP& Obfsproxy& 50& 0& 0& 0/0& 0& 0 \\
IVPN& TCP& Vanilla& 50& 50& 1& 50/50& 50& 1 \\
IVPN& UDP& Vanilla& 50& 50& 1& 50/50& 50& 1 \\
\rowcolor{myblue}
KeepSolid/Unlimited& TCP& Proprietory/XOR*& 50& 50& 1& 50/50& 50& 1 \\
\rowcolor{myblue}
KeepSolid/Unlimited& UDP& Proprietory/XOR*& 50& 50& 1& 50/50& 50& 1 \\
KeepSolid/Unlimited& TCP& Vanilla& 50& 50& 1& 50/50& 50& 1 \\
\rowcolor{myblue}
Mullvad& TCP& Shadowsocks& 50& 39& 0.78& 0/39& 0& 0 \\
Mullvad& TCP& Vanilla& 50& 47& 0.94& 47/47& 47& 0.94 \\
Mullvad& UDP& Vanilla& 50& 49& 0.98& 49/49& 49& 0.98 \\
\rowcolor{myblue}
NordVPN& TCP/UDP& XOR*& 50& 50& 1& 50/50& 50& 1 \\
NordVPN& TCP& Vanilla& 50& 50& 1& 48/50& 50& 1 \\
NordVPN& UDP& Vanilla& 50& 50& 1& 50/50& 50& 1 \\
\rowcolor{myblue}
PerfectPrivacy& TCP& Stunnel& 50& 47& 0.94& 0/47& 47& 0.94 \\
\rowcolor{myblue}
PerfectPrivacy& TCP& SSH& 50& 39& 0.78& 0/39& 39& 0.78 \\
\rowcolor{myblue}
PerfectPrivacy& TCP& Obfsproxy3& 50& 42& 0.84& 0/42& 42& 0.84 \\
PerfectPrivacy& TCP& Vanilla& 50& 50& 1& 50/50& 50& 1 \\
PerfectPrivacy& UDP& Vanilla& 50& 49& 0.98& 49/49& 49& 0.98 \\
\rowcolor{myblue}
PrivateInternetAccess& TCP& Shadowsocks& 50& 50& 1& 0/50& 50& 1 \\
PrivateInternetAccess& TCP& Vanilla& 50& 50& 1& 50/50& 50& 1 \\
PrivateInternetAccess& UDP& Vanilla& 50& 50& 1& 50/50& 50& 1 \\
\rowcolor{myblue}
PrivateVPN& TCP& Shadowsocks& 50& 50& 1& 0/50& 50& 1 \\
PrivateVPN& TCP& Vanilla& 50& 50& 1& 49/50& 50& 1 \\
PrivateVPN& UDP& Vanilla& 50& 50& 1& 50/49& 50& 1 \\
\rowcolor{myblue}
SecureVPN& TCP& SSH Tunnel& 50& 50& 1& 0/50& 50& 1 \\
\rowcolor{myblue}
SecureVPN& TCP& SSL Tunnel& 50& 49& 0.98& 0/49& 49& 0.98 \\
SecureVPN& TCP& Vanilla& 50& 50& 1& 50/50& 50& 1 \\
SecureVPN& UDP& Vanilla& 50& 49& 0.98& 48/49& 49& 0.98 \\
\rowcolor{myblue}
StrongVPN& TCP& XOR& 50& 50& 1& 50/50& 50& 1 \\
\rowcolor{myblue}
StrongVPN& UDP& XOR& 50& 50& 1& 50/50& 50& 1 \\
StrongVPN& TCP& Vanilla& 50& 49& 0.98& 49/49& 49& 0.98 \\
StrongVPN& UDP& Vanilla& 50& 49& 0.98& 49/49& 49& 0.98 \\
\rowcolor{myblue}
SurfShark& TCP& Proprietary/XOR*& 50& 50& 1& 50/50& 50& 1 \\
\rowcolor{myblue}
SurfShark& UDP& Proprietary/XOR*& 50& 50& 1& 50/50& 45& 0.9 \\
\rowcolor{myblue}
TorGuard& TCP& XOR*& 50& 50& 1& 49/50& 50& 1 \\
\rowcolor{myblue}
TorGuard& UDP& XOR*& 50& 50& 1& 50/50& 50& 1 \\
\rowcolor{myblue}
TorGuard& TCP& SSL Tunnel& 50& 44& 0.88& 0/44& 44& 0.88 \\
TorGuard& TCP& Vanilla& 50& 50& 1& 50/50& 50& 1 \\
TorGuard& UDP& Vanilla& 50& 49& 0.98& 49/48& 49& 0.98 \\
\rowcolor{myblue}
TunnelBear& TCP& Obfsproxy& 50& 0& 0& 0/0& 0& 0 \\
TunnelBear& TCP& Vanilla& 50& 50& 1& 50/50& 50& 1 \\
\rowcolor{myblue}
VPN.AC& TCP& XOR& 50& 50& 1& 50/50& 50& 1 \\
\rowcolor{myblue}
VPN.AC& UDP& XOR& 50& 50& 1& 50/50& 50& 1 \\
\rowcolor{myblue}
VPN.AC& TCP& V2Ray& 50& 0& 0& 0/0& 0& 0 \\
VPN.AC& TCP& Vanilla& 50& 50& 1& 50/50& 50& 1 \\
VPN.AC& UDP& Vanilla& 50& 50& 1& 50/50& 50& 1 \\
\rowcolor{myblue}
VPNArea& TCP& Stunnel& 50& 49& 0.98& 0/49& 49& 0.98 \\
VPNArea& TCP& Vanilla& 50& 50& 1& 50/50& 50& 1 \\
VPNArea& UDP& Vanilla& 50& 50& 1& 50/49& 50& 1 \\
\rowcolor{myblue}
VyprVPN& UDP& Proprietory& 50& 0& 0& 0/0& 0& 0 \\
VyprVPN& UDP& Vanilla& 50& 50& 1& 50/50& 0& 0 \\
\rowcolor{myblue}
Windscribe& TCP& TLS Tunnel& 50& 49& 0.98& 0/49& 25& 0.5 \\
\rowcolor{myblue}
Windscribe& TCP& WebSocket Tunnel& 50& 48& 0.96& 0/48& 24& 0.48 \\
Windscribe& TCP& Vanilla& 50& 50& 1& 50/50& 50& 1 \\
Windscribe& UDP& Vanilla& 50& 50& 1& 50/50& 50& 1 \\
       	
    \bottomrule
    \end{tabular*}
    \caption{\textbf{Evaluation results on \Merit, breakdown by configuration.} Highlighted rows are ``obfuscated'' configurations. Variants marked with stars mean that the VPN provider does not disclose which obfuscation technique is used and we can only infer the variant type based on packet captures. Note Hide.me claims the \textit{tls-crypt} option alone is enough to ``obfuscate entire traffic''~\cite{hide-me-obfs}. However, this option only encrypts control channel payloads but not the OpenVPN packer headers.}
    \label{tab:merit-results}
\end{table*}

\end{document}